\documentclass[11pt,a4paper]{article}
\usepackage[utf8]{inputenc}
\usepackage[T1]{fontenc}
\usepackage{lmodern}
\usepackage{minibox}

\usepackage[left=2.5cm,top=2.5cm,right=2.5cm,bottom=2.5cm]{geometry}
\usepackage{setspace}
\usepackage{amsmath,amssymb,amsfonts,amsthm,bm,latexsym}
\usepackage{nicefrac}

\usepackage{array}
\usepackage{booktabs}
\usepackage{longtable}
\usepackage{multirow}
\usepackage{float}
\usepackage{rotating}
\usepackage{pdflscape}
\usepackage{morefloats}
\usepackage{placeins}

\usepackage{enumitem}

\newcolumntype{C}[1]{>{\centering\arraybackslash}m{#1}}

\renewcommand{\arraystretch}{1.03}
\usepackage{graphicx}
\usepackage{svg}

\usepackage{caption}
\usepackage[authoryear]{natbib}
\usepackage[bottom]{footmisc}
\usepackage{xurl}
\usepackage{microtype}
\usepackage[none]{hyphenat}

\usepackage{color}
\usepackage{soul}
\usepackage{todonotes}
\usepackage{comment}

\usepackage{authblk}
\usepackage{multicol}
\usepackage{pifont}

\usepackage{xr-hyper}
\usepackage{hyperref}
\hypersetup{
    colorlinks=true,
    breaklinks=true,
    linkcolor=[rgb]{0,0.3,1},
    citecolor=[rgb]{0,0.3,1},
    urlcolor=[rgb]{0,0.3,1}
}

\newtheorem{theorem}{Theorem}

\newtheorem{claim}{Claim}
\newenvironment{cl}{\vskip.2cm\begin{claim}}{\end{claim}\vskip.2cm}

\newtheorem{lemma}{Lemma}

\newtheorem{corolla}{Corollary}

\newtheorem{defini}{Definition}
\newenvironment{defi}{\vskip.2cm\begin{defini}}{\end{defini}\vskip.2cm}

\newtheorem{proposi}{Proposition}

\newtheorem{prop}{Proposition}

\newtheorem{cla}{Claim}

\newtheorem{assump}{Assumption}

\newtheorem{demo}{Proof}

\newtheorem{remark}{Remark}

\newtheorem{example}{Example}

\newtheorem{notat}{Notation}

\def\cqfd {$\Box$}
\def\co{{\rm co}\hskip 1pt}

\def\E{{\mathcal E}}

\def\F{{\mathcal F}}

\def\H{{\mathcal H}}

\def\M{{\mathcal M}}

\def\proj{{\rm proj}}
\def\proj0{{\rm proj}_{\co M_0}}

\def\R{{\mathbb{R}}}

\begin{document}
\title{Structural economic change under the digital transformation}

\author[1,3,*]{Marco Due\~nas}
\author[2,3]{Antoine Mandel}
\affil[1]{\small Institute of Economics -- Sant'Anna School of Advanced Studies, Pisa, Italy}
\affil[2]{\small Centre d'Economie de la Sorbonne --  Paris School of Economics -- CNRS-Universit\'{e} Paris 1 Panth\'{e}on-Sorbonne, Paris, France}
\affil[3]{\small CLIMAFIN -- Climate Finance Alpha, Paris, France}
\affil[*]{Corresponding author email: marco.duenas@climafin.com}

\date{}

\maketitle
\vspace{-40pt}

\begin{abstract}
\noindent Productivity improvements in digital industries could massively reshape the structure of production networks and drive significant labour reallocation across sectors. We examine these adjustments using a calibrated agent-based model of the global economy. We consider two scenarios with the same cumulative increase in digital-supplier productivity, which differ in how digital inputs in exposed industries are incorporated into production. The key difference is whether digital inputs enter as standard intermediate inputs (low embeddedness) or are combined directly with primary factors (high embeddedness). Across economies, both scenarios generate gains in gross output, household utility, and GDP, although value-added gains are more modest and unevenly distributed across sectors. High embeddedness generally amplifies production responses, particularly in industries highly exposed to digital inputs. Structurally, as digital industries become central in the production network and attract labour from other activities, other industries move further upstream and downstream. Persistent labour reallocation pressures also indicate potential unemployment risks where workers cannot move to expanding industries. These findings stress that, beyond productivity gains derived from advances in digital technologies, economic and social consequences depend on how digital inputs integrate into production and on workers' possibilities to adapt to structural change.
\end{abstract}


\medskip

\clearpage
\newpage
\section{Introduction}
\label{sec:introduction}

Digital technologies are transforming how production and labour are organised, with major social, political, and economic consequences \citep{WorldBank2016Digital, GmyrekEtAl2023, OECD2024Digital}. Beyond generating productivity gains, these technologies can alter how firms combine productive factors, interact with their suppliers, and demand labour \citep{bresnahan2002information, GoldfarbTucker2019}. From this perspective, digital transformation represents a process of structural change that extends beyond an additional source of productivity growth. Therefore, understanding its economic implications requires examining how technological progress modifies inter-industry interdependencies and labour reallocation requirements.

Recent literature has investigated the impacts of digital technologies on productivity and labour demand \citep{AcemogluRestrepo2019, DauthEtAl2021}, as well as the modifications they induce in complementarity and substitutability among productive factors \citep{bresnahan2002information, SchwarkTryphonides2025}. Their broader structural impacts, however, depend on the propagation of these productivity shocks across industries \citep{carvalho2019production, baqaee2019macroeconomic}. This paper investigates these structural issues by analysing how the embedding of digital inputs into production shapes the recomposition of production networks and the sectoral reallocation of labour.

We employ an agent-based model of economic dynamics in production networks calibrated using EXIOBASE \citep{stadler2018exiobase}. The model incorporates industries interconnected through input-output relations, gradual adjustment of prices, production, and expenditure decisions \citep{gualdi2016emergence, mandel2021monetary}, and inter-sectoral labour mobility \citep{duenas2026labour}. In this setting, we analyse the structural impacts induced by productivity shocks in digital industries in two scenarios with alternative assumptions about the depth at which digital inputs are embedded in production technologies. Both scenarios rely on the identification of digitally exposed industries \citep{CalvinoEtAl2018, CalvinoEtAl2024, PrytkovaEtAl2025}. Under low embeddedness, digital inputs, for both exposed and non-exposed industries, enter production as standard intermediate inputs, with substitution against capital and labour occurring through the upper-level production aggregator. Under high embeddedness, only digitally exposed industries combine digital inputs directly with capital and labour, allowing greater substitution with primary factors.  
We consider the same exogenous productivity shock on digital industries in both scenarios and investigate the differentiated propagation mechanisms and structural effects implied by the specific technological configurations in each scenario.

Our results highlight three interconnected dimensions of the digital transition: aggregate economic performance, the recomposition of production networks, and labour reallocation. Across both scenarios, digital productivity gains drive heterogeneous economic expansion across countries and industries. At the aggregate level, the transition lowers prices while increasing gross output, GDP, and household utility. However, this expansion is more pronounced in gross output than in GDP. Similarly, at the sectoral level, gross output responds more strongly than value added, whose gains are generally more modest and even negative in certain country--industries. This suggests that the economic dynamism triggered by digital productivity improvements across production chains does not necessarily translate into proportional gains in value added. These aggregate results are qualitatively robust to variations in the magnitude of the digital productivity shock and in substitution elasticities within the examined ranges.

Both scenarios also induce a structural recomposition of production networks. Digital industries reduce their intermediate input requirements per unit of output while increasing their prominence as direct and indirect suppliers to other activities. This growing importance as suppliers is accompanied by a substantial expansion of digital production and value added. Digital industries also emerge as a major destination of labour reallocation, with services constituting an important source of workers in several economies. However, the origins and destinations of labour flows vary across countries, depending on their productive structures and mobility constraints. Greater labour mobility reduces aggregate reallocation gaps, although residual pressures persist across countries and skill groups.

While both scenarios generate broadly similar patterns of structural adjustment, high embeddedness generally amplifies economy-wide production responses, particularly in highly exposed industries. Greater integration of digital inputs with capital and labour also reinforces the importance of digital industries as suppliers to other activities, while moderating the decline in their own intermediate-input requirements per unit of output. Regarding their positions within production chains, digital industries become more upstream and less downstream relative to the benchmark, whereas other economic sectors generally show increases in both upstreamness and downstreamness. Labour adjustments also remain heterogeneous: under high embeddedness, for instance, manufacturing in China and Korea can experience increases in allocated working hours for some skill groups, while other sectors generally lose labour. 

Taken together, these findings emphasise an important tension with significant economic and social implications: digital productivity improvements can generate aggregate economic gains while simultaneously creating uneven adjustment pressures across industries and labour markets.

Our analysis has a deliberately conditional scope. We do not model the endogenous growth of primary factor endowments, technology adoption, or autonomous productivity shifts in sectors other than digital suppliers. Consequently, our objective is not to project a future trajectory of digital transformation, but to construct a counterfactual exercise that shows how existing productive structures might respond to a substantial improvement in digital productivity.

The remainder of the paper is organised as follows. Section~\ref{sec:literature} reviews the related literature. Section~\ref{sec:model} presents the economic model, alternative technological structures, and adjustment mechanisms. Section~\ref{sec:results} presents the simulation results, and Section~\ref{sec:concluding_remarks} concludes. Supplementary Material (SM) documents additional details on the data, model specification and implementation, and auxiliary results.

\section{Related literature}
\label{sec:literature}
Digital technologies can reshape production, trade, and labour markets by reducing the costs of processing, transmitting, and verifying information \citep{GoldfarbTucker2019}. The economic consequences have been examined at different levels of analysis. 

At the microeconomic level, research has focused on how new technologies affect tasks, firm productivity, production organisation, and labour demand. At the macroeconomic level, studies have examined how technological improvements translate into aggregate productivity, output, and welfare. While these perspectives illuminate different dimensions of technological change, an important question is how productivity improvements originating in particular industries translate into broader economic and structural adjustments. Production-network analysis provides a framework for addressing this question by tracing how shocks propagate through supplier--customer linkages, affecting both aggregate economic performance and the organisation of production.

In the context of digital transformation, an important distinction arises from the heterogeneity of digital technologies and their uses. The OECD Digital Intensity and AI taxonomies distinguish several dimensions of digitalisation, technological exposure, and actual use \citep{CalvinoEtAl2018, CalvinoEtAl2024}. This heterogeneity motivates a distinction between industries that \emph{supply} digital goods and services and those potentially \emph{exposed} to their use. Exposure identifies where digital technologies may matter, but does not measure realised adoption or establish how digital inputs substitute for other factors.

The consequences for employment are similarly heterogeneous. Task-based models distinguish the displacement of labour from tasks that become automated from productivity and reinstatement effects that can support labour demand \citep{AcemogluRestrepo2019}. Empirical studies report different outcomes across technologies and labour markets: robot exposure has reduced employment and wages in some US local markets \citep{AcemogluRestrepo2020}, whereas manufacturing job displacement associated with robots in Germany was offset by employment growth in services \citep{DauthEtAl2021}. Patent-based evidence on emerging digital technologies also reveals differences across technologies and skill groups \citep{PrytkovaEtAl2025}. These findings motivate examining sectoral changes in labour allocation and remuneration, rather than inferring employment outcomes from aggregate productivity gains. Our framework follows this intersectoral margin without modelling the automation or creation of individual tasks.

A related question is how digital technologies alter the combination of productive inputs. \citet{bresnahan2002information} document complementarities between information technology, workplace organisation, and skilled labour, emphasising that technological benefits depend on the resources with which digital technologies are combined. More directly, \citet{SchwarkTryphonides2025} find that digital capital intensity is associated with changes in substitution elasticities, particularly between capital and labour and between value added and intermediate inputs. These effects differ across IT, software and databases, and communication equipment. Their evidence suggests that digitalisation can affect production flexibility as well as productivity. Still, it does not identify a common elasticity or a unique production-function placement for \emph{purchased} digital inputs.

The economic effects of digital technologies also depend on how digital inputs interact with capital and labour in production. Related work, such as \citet{melina2025From}, examines how alternative substitution relationships involving ICT capital condition macroeconomic responses. This motivates our comparison of alternative production structures through which digital productivity improvements propagate across industries.

Production-network theory explains why this technological distinction can matter beyond the firms directly using digital inputs. Productivity improvements propagate through input-output linkages, while changes in relative prices and input combinations can redirect demand across interconnected industries \citep{carvalho2019production}. \citet{baqaee2019macroeconomic} show that the effects of sectoral productivity shocks beyond first-order approximations depend on substitution elasticities, production linkages, and factor reallocation. Thus, even when alternative technologies reproduce the same initial transactions, their responses to a common shock may differ as input requirements and expenditure weights adjust. These results motivate examining both aggregate outcomes and the changing intensity and centrality of production linkages. Unlike models of endogenous supplier formation \citep{oberfield2018theory, Acemoglu2020Endogenous}, our simulations keep feasible trading relationships fixed and allow their weights to evolve. These theoretical results motivate our analysis, although they do not directly describe the disequilibrium adjustment processes simulated in our model.

The connection to labour markets follows from these production adjustments. Changes in the composition of intermediate demand can redistribute activity and labour requirements across industries, including those that do not receive a productivity shock directly. Evidence from global value chains highlights the distinct roles of labour productivity, input--output linkages, and final demand in employment changes \citep{BontadiniEtAl2026}. Thus, we investigate the joint adjustment of sectoral production, wages, and skill-specific working hours, allowing labour mobility to respond gradually to remuneration differences and technological proximity between industries. With national skill-specific labour endowments fixed, our outcomes concern intersectoral reallocation and remaining equal-remuneration gaps, not changes in aggregate employment or observed vacancies.

Finally, the future scale of digital productivity gains and their diffusion across the economy remain uncertain. Assessments of AI's macroeconomic effects depend on assumptions about technological capabilities, adoption, and the translation of innovation into effective productivity \citep{FilippucciGalSchief2024}; multisector digital-investment scenarios likewise depend on assumed transmission channels \citep{CasasEtAl2026}. We therefore introduce a common exogenous total factor productivity (TFP) improvement in \emph{digital industries}, representing their ability to produce more output from a given input bundle, rather than imposing productivity gains directly on every exposed user.

Our contribution is to connect these strands in a multi-country monetary input--output framework with gradual disequilibrium adjustment and frictional labour mobility \citep{gualdi2016emergence, mandel2021monetary, duenas2026labour}. We examine how integrating digital inputs conditions the propagation of supplier-side productivity gains, the recomposition of weighted production networks, and the sectoral distribution of labour and remuneration. This approach links technological uncertainty to simulated economic outcomes while distinguishing aggregate gains from the adjustment pressures that can persist during structural transformation.

\section{Economic model}
\label{sec:model}
We investigate how digital productivity improvements propagate using an agent-based model of economic dynamics in production networks that simulates out-of-equilibrium dynamics in a general-equilibrium framework \citep{gualdi2016emergence,mandel2021monetary}. Firms and households interact through interconnected production and expenditure networks, while prices, production, inventories, expenditure decisions, and monetary balances adjust gradually rather than through instantaneous equilibrium adjustment. The model also incorporates frictional labour mobility across industries, allowing technological shocks to reshape both production linkages and sectoral labour allocation \citep{duenas2026labour}. Below we summarise the main economic mechanisms, leaving the complete formal specification and updating rules to Section~\ref{SM-sec:model_SM}.

The model is calibrated to the 2022 world economy using EXIOBASE Version 3.10.1 \citep{stadler2018exiobase}, covering $L=163$ industries across $C=49$ countries or regions: the twenty-seven EU member states, seventeen other major economies, and five aggregate rest-of-the-world regions.\footnote{2022 is the latest available data for supply-use tables and bilateral trade data.} Each country--industry pair is represented by a \textit{firm} supplying a differentiated product. Within each country, three representative worker households supply low-, medium-, and high-skill labour, respectively, while a fourth household supplies capital services. National labour and capital endowments are fixed, but labour can be reallocated across domestic industries, and firms source intermediate inputs from domestic and foreign producers. The resulting economy combines national factor markets with international production networks.

\subsection{General equilibrium setup}

Let $\F$ denote the set of firms and $\H=\H_L\cup\H_K$ the set of households, comprising workers ($\H_L$) and capital owners ($\H_K$). Firms use nested constant-elasticity-of-substitution (CES) production technologies that combine domestic capital, skill-specific labour, and intermediate inputs from domestic and foreign suppliers. Capital and the three labour types form a value-added nest, while intermediate inputs are organised into separate nests according to their characteristics. Likewise, households have nested CES preferences over consumption goods. The nesting structure and substitution elasticities determine how firms and households adjust their input and consumption choices to changes in relative prices, providing an important channel through which productivity improvements propagate across industries. The complete classification of production and consumption nests is reported in Section~\ref{SM-sec:digital_class}.

The initial state of the economy is calibrated to a general equilibrium that reproduces the input--output transactions, production levels, and factor payments reported in EXIOBASE. At this equilibrium, firms' expenditures on intermediate inputs and primary factors are consistent with their production technologies, while labour and capital remuneration constitute household income. These transactions define an expenditure-share matrix describing how each agent's spending is distributed across suppliers, providing the initial structure through which changes in production costs, prices, and demand propagate among interconnected industries \citep{baqaee2019macroeconomic, carvalho2019production}. As the economy adjusts to technological changes, expenditure shares and production relationships evolve endogenously. The formal equilibrium conditions and calibration procedures are provided in Section~\ref{SM-sec:S_model_calibration}.

\subsubsection{Labour market}

At the benchmark equilibrium, wages are equal across firms within each country and skill group, so no wage-driven reallocation occurs initially. Following a productivity shock, differences in sectoral activity and labour requirements generate wage differentials, creating incentives for workers to move towards better-paying firms.

Labour mobility is also constrained by technological proximity between firms, which serves as a proxy for the \emph{transferability of skills}. We measure this proximity using the cosine similarity between firms' input-coefficient vectors, corresponding to the columns of the initial technical-coefficient matrix $\mathbf{Z}^*=(z_{ij,t=0})_{i,j\in\F}$.

Therefore, given a productivity shock in the production network, labour flows depend on wage differentials and the transferability of skills and, consistent with the out-of-equilibrium dynamics, adjust gradually, governed by a mobility parameter $\tau_h$. This leads to remuneration differences that may persist during the transition.

To assess the remaining pressures for labour reallocation, we compare allocated working hours with a reference allocation that would equalise hourly remuneration across industries within each country and skill group. Let $\ell^t_{h,i}$ and $p^t_{h,i}$ denote working hours and wages for skill group $h$ in industry $i$, and $e_h$ its fixed national labour endowment. Defining the hours-weighted average wage as $\tilde{p}_h^t=\sum_i p^t_{h,i}\ell^t_{h,i}/e_h$, the reference allocation and labour gap are
\begin{equation}
\label{eq:labour_gap}
\overline{\ell}^{\,t}_{h,i}
=\ell^t_{h,i}\frac{p^t_{h,i}}{\tilde{p}_h^t},
\qquad
\Delta\ell^t_{h,i}
=\ell^t_{h,i}-\overline{\ell}^{\,t}_{h,i}=
\ell^t_{h,i}\left(1-\frac{p^t_{h,i}}{\tilde{p}_h^t}\right).
\end{equation}
Intuitively, industries offering above-average wages would receive working hours, generating negative gaps and potential pressures for labour inflows. Conversely, industries with below-average wages would lose working hours, generating positive gaps and potential pressures for labour outflows. These gaps capture potential reallocation pressures rather than observed vacancies or unemployment; therefore, the reference allocation is a diagnostic tool and does not represent a new market-clearing equilibrium.

\subsection{Digital transformation: nesting and productivity}
\label{sec:digital_calibration}
We distinguish between \emph{digital industries}, which produce digital goods and services, and \emph{digitally exposed industries}, whose production technologies may allow these inputs to interact more closely with capital and labour. Digital suppliers are identified from the EXIOBASE classification and include activities related to computing, telecommunications, electronics, and other ICT goods and services. Their products constitute a distinct category of intermediate inputs, hereafter referred to as \emph{digital inputs}. Table~\ref{SM-tab:S_industries} reports the corresponding classification.

Our key focus is the impact of productivity growth in digital industries on industries exposed to substitution/competition from digital inputs. To identify potentially exposed industries, we draw on the OECD Digital Intensity taxonomy \citep{CalvinoEtAl2018} and the OECD AI taxonomy \citep{CalvinoEtAl2024}. Both classify industries into quartiles (Q1--Q4), with higher quartiles indicating greater digital intensity or AI exposure. Industries identified in the upper quartiles (Q3--Q4) are classified as exposed, using the TechXposure indicator \citep{PrytkovaEtAl2025} as an additional measure to validate and refine the mapping.\footnote{These indicators provide an empirical basis for identifying potential technological exposure, but do not determine the substitution elasticities of the production function.} This procedure identifies 49 exposed industries among the 163 EXIOBASE industries. Table~\ref{SM-tab:S_industries} reports the complete classification.

In our modelling setup, digital industries face exogenous TFP improvements, while productivity in all other industries remains unchanged. The simulations begin with a one-year pre-shock period, corresponding to $t'=\Delta t$ iterations, during which productivity remains at its benchmark level. Thereafter, digital-supplier TFP increases progressively over $T_g$ iterations and remains constant for the rest of the simulation. Formally, let $\mathcal D$ denote the set of all producers belonging to the digital industry. The TFP evolution for any firm $i \in \F$ is defined as:
\begin{equation}
\Lambda_{i,t}=
\begin{cases}
1, & i\notin\mathcal D \text{ or } t\leq t',\\
(1+g_D)^{(t-t')/T_g}, & i\in\mathcal D,\quad t'<t\leq t'+T_g,\\
1+g_D, & i\in\mathcal D,\quad t>t'+T_g,
\end{cases}
\label{eq:digital_tfp}
\end{equation}
where $g_D$ denotes the cumulative TFP increase over the innovation period $T_g$ in the digital industry. In the benchmark, $g_D=0.35$ and $T_g=200$ iterations, corresponding to ten years of compounded productivity growth following the initial pre-shock year. Productivity then remains constant for an additional five years of simulation.\footnote{Benchmark outcomes typically stabilise within approximately one year (20 iterations) after productivity growth ends. However, some alternative parametrisations, particularly those involving higher substitution elasticities
or larger productivity shocks can induce more substantial adjustments across production networks, requiring longer adjustment periods.
We therefore retain a common five-year post-shock period across
all simulations to ensure comparability. This choice does not affect the benchmark results.}

We now present the technological configurations. Digital inputs can be identified in observed input--output transactions; however, their technological relationship with capital and labour is less directly observable. Within the nested CES framework, this relationship depends on how digital inputs are combined with primary factors. 

We compare two production structures, both calibrated to the same reference economy and subject to an identical digital productivity shock, to examine how substitution possibilities affect the propagation of digital productivity shocks.
\begin{itemize}
\setlength{\itemsep}{2pt}
    \item Low-embeddedness scenario (Low). Digital inputs remain in a separate intermediate-input nest in all industries. Substitution with capital and labour occurs through the upper-level CES aggregator.
    \item High-embeddedness scenario (High). In the 49 exposed industries, digital inputs enter an augmented capital--labour nest, allowing substitution with primary factors at the within-nest elasticity. The remaining industries retain the low-embeddedness production structure.
\end{itemize}
Formally, let $\mathcal E$ denote the set of exposed firms, and $V_i$ the domestic capital and skill-specific labour inputs of firm $i \in \F$. The corresponding input partitions are
\begin{equation}
S_i^{\mathrm{Low}}=\{V_i,\mathcal D,\ldots\},\qquad
S_i^{\mathrm{High}}=
\begin{cases}
\{V_i\cup\mathcal D,\ldots\}, & i\in\mathcal E,\\
S_i^{\mathrm{Low}}, & i\notin\mathcal E,
\end{cases}
\label{eq:digital_partition}
\end{equation}
where the remaining intermediate-input nests are unchanged. These alternative partitions modify substitution possibilities without introducing new supplier links. Under the benchmark parameters in Table~\ref{SM-tb:sim_parameters}, digital inputs and primary factors share a within-nest substitution elasticity of $0.80$ in exposed industries under the high-embeddedness scenario, compared with an upper-level elasticity of $0.33$.\footnote{Including digital inputs in the capital--labour nest does not change their accounting treatment as intermediate consumption when calculating gross value added.}

\subsection{Agent-based dynamics}

Starting from the calibrated initial benchmark equilibrium ($t_0$), the model simulates the economy's response to exogenous productivity improvements in digital industries. 

In each period, agents allocate their monetary balances across suppliers according to their expenditure shares, generating demand for goods and factor services. Prices and wages adjust gradually towards market-clearing values, while available production and inventories determine how much demand can be satisfied. Firms produce using the inputs they obtain, and revenues and expenditures update agents' balances. Firms also adjust their input combinations and households their consumption patterns in response to relative prices, with expenditure shares adapting gradually. Labour is reallocated across domestic industries according to wage differentials and technological proximity, subject to the mobility parameter $\tau_h$.

Productivity growth in digital industries changes other industries' input choices and production possibilities, and the resulting shifts in sectoral activity feed back into labour demand, remuneration, and household expenditure. Because adjustment is frictional, the transition may involve persistent imbalances as network weights evolve endogenously.

The CES parameters are calibrated separately for each scenario to reproduce the same initial input--output observations (Section~\ref{SM-sec:S_model_calibration}). The outcomes reported at the end of the simulation ($T$) therefore describe the economy after a finite adjustment horizon---not necessarily a new equilibrium.

\section{Results}
\label{sec:results}
We present the results for the low- and high-embeddedness scenarios for a selected set of countries, with the EU-27 reported separately as an aggregate. The benchmark simulation results use a cumulative 35\% increase in digital-supplier TFP, a value-added substitution elasticity of $\sigma_{\mathrm{VA}}=0.80$, and frictional labour mobility ($\tau_h=0.015$). We use this benchmark to present results on aggregate economic performance, the recomposition of production networks, and labour reallocation. Later, we extend our analysis to understand labour adjustment pressures considering different mobility regimes, while the final subsection presents a sensitivity analysis to alternative productivity shocks and substitution elasticities. We focus on outcomes at the end of the simulation ($T$), rather than on adjustment dynamics, and report most results relative to their calibrated $t_0$ reference values.

\subsection{Aggregate prices, output, and GDP}
\begin{figure}[h!]
\centering
\includegraphics[width=0.99\linewidth]{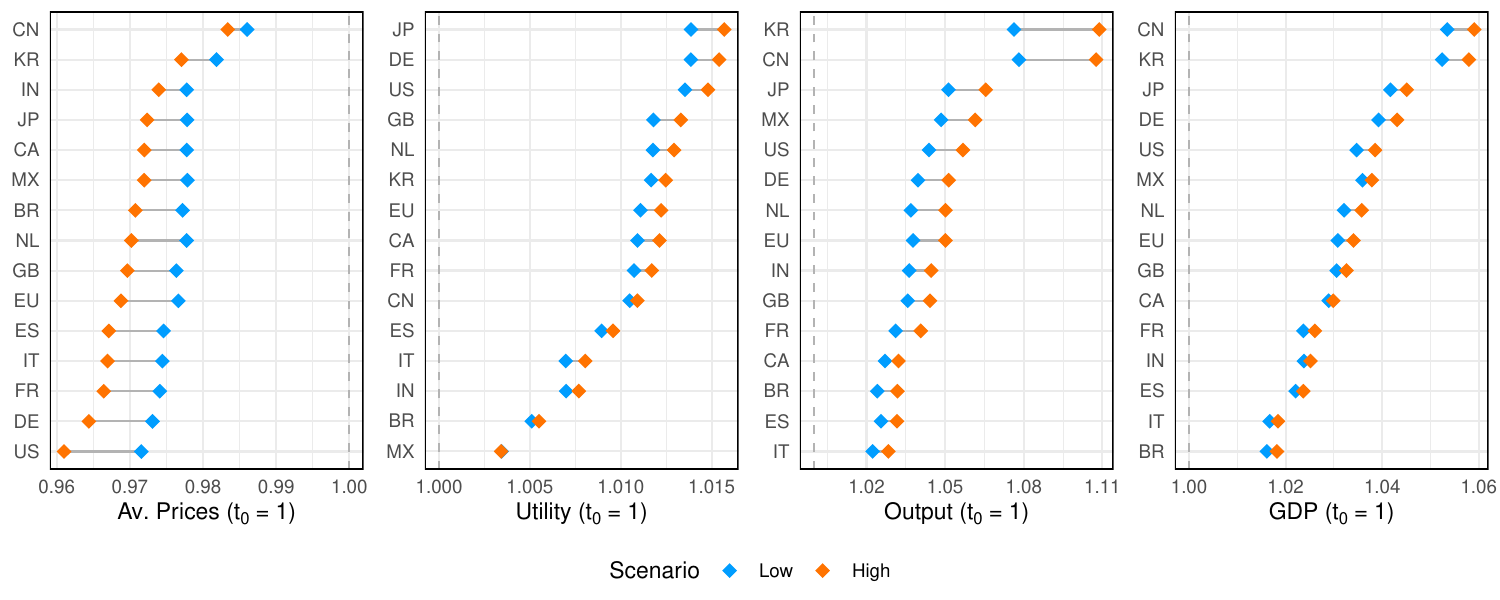}
\caption{Aggregate outcomes at the end of the simulation $T$, relative to the initial benchmark ($t_0=1$). Panels from left to right report average prices, aggregate household utility, gross output, and GDP. Colours distinguish low- and high-embeddedness scenarios.}
\label{fig:d_gva}
\end{figure}

Turning first to macroeconomic aggregates, Figure~\ref{fig:d_gva} shows a common pattern across all economies: productivity improvements in digital industries lead to lower average prices and higher gross output, GDP, and household utility in both scenarios. High embeddedness strengthens all these responses. For the EU aggregate, gross output is approximately 4\% above its initial level under low embeddedness and 5\% under high embeddedness. GDP gains are around 3\% in both scenarios, with a slightly larger increase under high embeddedness. Korea and China exhibit the largest gross output responses, reaching approximately 11\% under high embeddedness. China records the largest GDP increase, close to 6\%, followed closely by Korea. The largest price decline occurs in the United States, approaching 4\% under high embeddedness. We also report changes in household utility, which increases in every economy considered. However, this aggregate measure does not reveal how these gains are distributed across different household types.

Because both embeddedness scenarios follow the same productivity path in digital industries, differences in economic outcomes arise from alternative production structures. Under high embeddedness, digital inputs enter the capital--labour nest of exposed industries, allowing firms greater substitution possibilities between digital inputs and primary factors as relative prices change. These adjustments affect intermediate-input demand and propagate through domestic and international production networks, influencing production and demand across sectors.

At the macroeconomic level, the digital transition leads to increases in gross output that exceed those in GDP. Since gross output includes intermediate production throughout the supply chain, whereas GDP measures value added, the larger gross-output response does not translate into equally large increases in aggregate value added or household utility. This distinction is particularly relevant when comparing technological scenarios: under high embeddedness, gross production is further amplified, so greater integration of digital inputs into production can substantially expand intermediate production without generating proportional increases in aggregate value added. The sectoral results below provide further insight into these differences.

\subsection{Sectoral output and value added}
In what follows, countries and the EU-27 aggregate are ordered by the digital-sector gross-output response under the high-embeddedness scenario.

\begin{figure}[h!]
\centering
\includegraphics[width=0.99\linewidth]{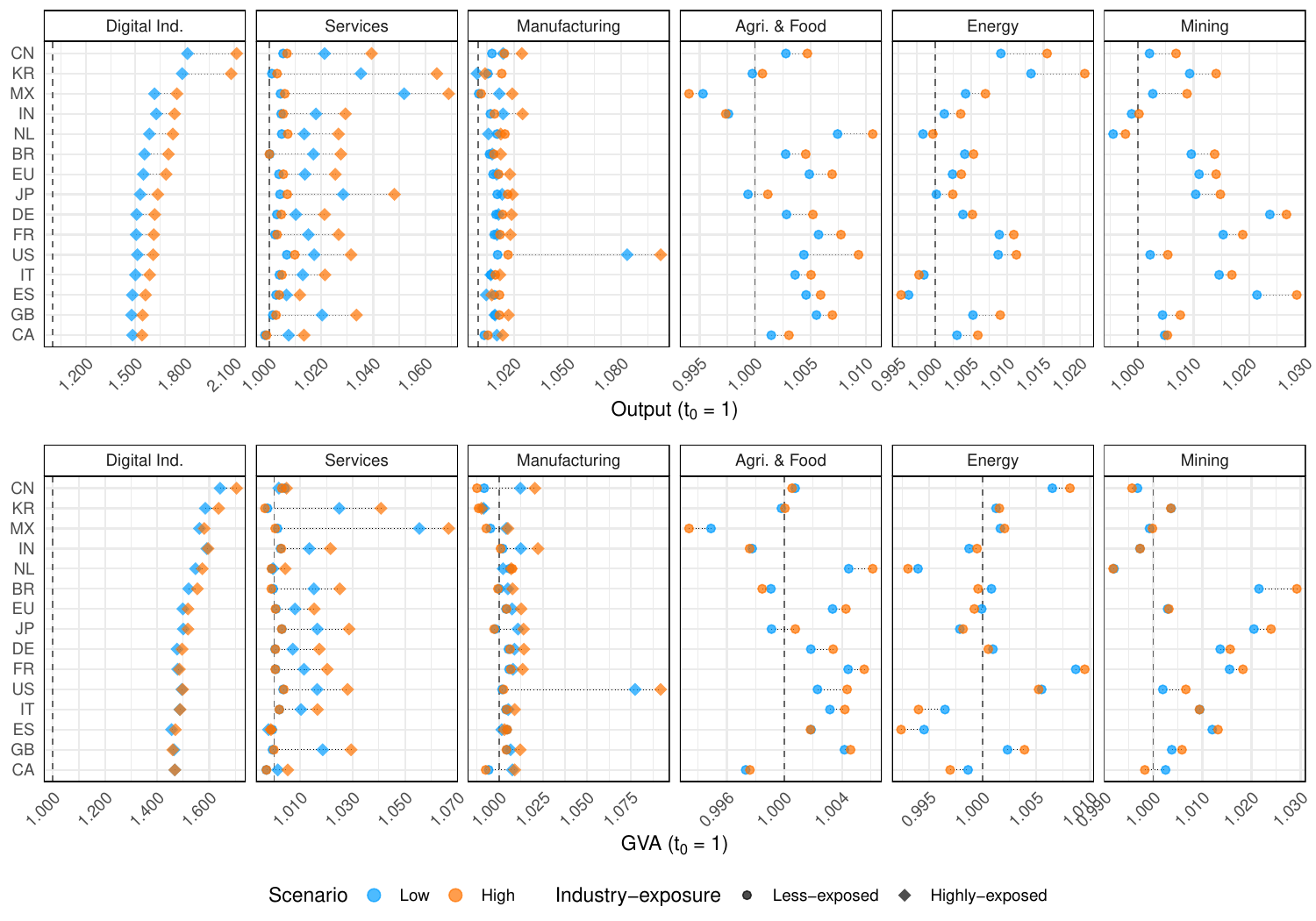}
\caption{Sectoral gross output (upper panel) and GVA (lower panel) at the end of the simulation, relative to the initial benchmark ($t_0=1$). Colours distinguish the Low- and High-embeddedness scenarios, while symbols distinguish less-exposed and highly-exposed industries to digital inputs.}
\label{fig:d_output_gva}
\end{figure}

The upper panel in Figure~\ref{fig:d_output_gva} reveals a highly uneven sectoral response to digital productivity improvements. Digital industries experience by far the largest expansion in gross output, with China and Korea reaching approximately 2.0--2.1 times their initial production levels under the high-embeddedness scenario. Substantial growth also occurs under low embeddedness, as both scenarios benefit from the same productivity improvement in digital industries. The stronger response under high embeddedness is consistent with greater substitution possibilities between digital inputs and primary factors in exposed industries, which can increase demand for digital inputs and stimulate production in digital industries.

Outside digital industries, output responses are considerably smaller and vary across sectors and exposure groups. Highly-exposed industries generally experience stronger expansions than less-exposed industries, particularly in services. Interestingly, this pattern is already visible under the low-embeddedness scenario, suggesting that industries classified as highly exposed can benefit from digital productivity improvements through existing production linkages, even without the greater substitution possibilities introduced under high embeddedness. The high-embeddedness scenario further amplifies these responses, particularly in Korea and Mexico.

Manufacturing exhibits a more heterogeneous pattern. Highly-exposed manufacturing industries in the United States stand out, with gross output increasing by approximately 10\% under high embeddedness, while responses in other economies are considerably smaller. In contrast, less-exposed manufacturing industries generally experience modest changes. Energy and mining show both gains and contractions, while agriculture and food remain below their initial output levels in Mexico and India under both scenarios. Thus, the large expansion of digital industries coexists with uneven responses across other sectors and countries.

The contrast between gross output and GVA is particularly revealing in the lower panel of Figure~\ref{fig:d_output_gva}. Digital-sector GVA increases substantially, with gains ranging from approximately 48\% in the United Kingdom to just over 70\% in China under high embeddedness. Outside digital industries, changes in GVA are generally smaller than those in gross output, although important differences emerge between exposure groups. Highly-exposed services experience notable gains in countries such as Mexico and Korea, while highly-exposed manufacturing in the United States records an increase of approximately 9\%. In contrast, GVA in less-exposed manufacturing remains close to its initial level or contracts in several economies. Agriculture and food GVA contracts in Mexico, India, and Canada despite higher aggregate GDP.

These differences also reflect the accounting distinction between gross output and value added. GVA deducts digital purchases as intermediate consumption even when those inputs enter the augmented value-added production nest; changing the technological partition does not change this accounting treatment.

High embeddedness generally amplifies the expansion of digital industries and highly-exposed activities, but does not systematically raise GVA across all sectors. Some activities experience smaller gains or larger contractions, as illustrated by agriculture and food in Mexico and mining in Canada. Overall, the results show that the benefits of digital productivity improvements depend on both the initial production structure and the technological possibilities for integrating digital inputs with primary factors.

\subsection{Production-network recomposition}
To investigate how digital productivity improvements reshape interindustry production linkages, we consider the matrix of real technical coefficients $\mathbf Z_t=(z_{ij,t})$, where $z_{ij,t}=x_{ij,t}/y_{j,t}$ denotes the quantity of intermediate input $i$ required per unit of gross output in industry $j$. Input quantities and gross output are expressed in constant $t_0$ prices. Unlike nominal expenditure shares, technical coefficients capture changes in the composition and intensity of intermediate input use without the mechanical revaluation caused by changes in current prices. They can nevertheless respond endogenously to relative-price changes. Since the set of feasible production linkages remains fixed, network recomposition takes place through changes in the weights of existing input--output relationships. 

To characterise the recomposition of production networks, we use indicators that distinguish an industry's dependence on intermediate inputs from its importance as a supplier (see Section~\ref{SM-sec:S_network_measures} for definitions). Specifically, we focus on in-strength, which measures the intermediate inputs required per unit of an industry's gross output, and out-strength, which captures the importance of its products in other industries' input requirements. These two dimensions need not move together: an industry may become less dependent on intermediate inputs in its own production while becoming a more important supplier to other activities. We also report the backward and forward Katz centralities, which extend in- and out-strength measures to indirect production relationships, capturing how an industry depends on suppliers or contributes to production through successive connections across the production network.

We also examine upstreamness and downstreamness \citep{antras2012measuring, bartolucci2025upstreamness}, which describe industries' positions along production chains rather than the intensity of their linkages. Upstreamness measures the average distance of an industry's output from final demand: a higher value indicates that its products pass through more production stages before reaching final consumers. Conversely, downstreamness measures the average distance from primary factors, reflecting the stages of intermediate production embodied in an industry's own inputs. 

\begin{figure}[h!]
\centering
\includegraphics[width=0.99\linewidth]{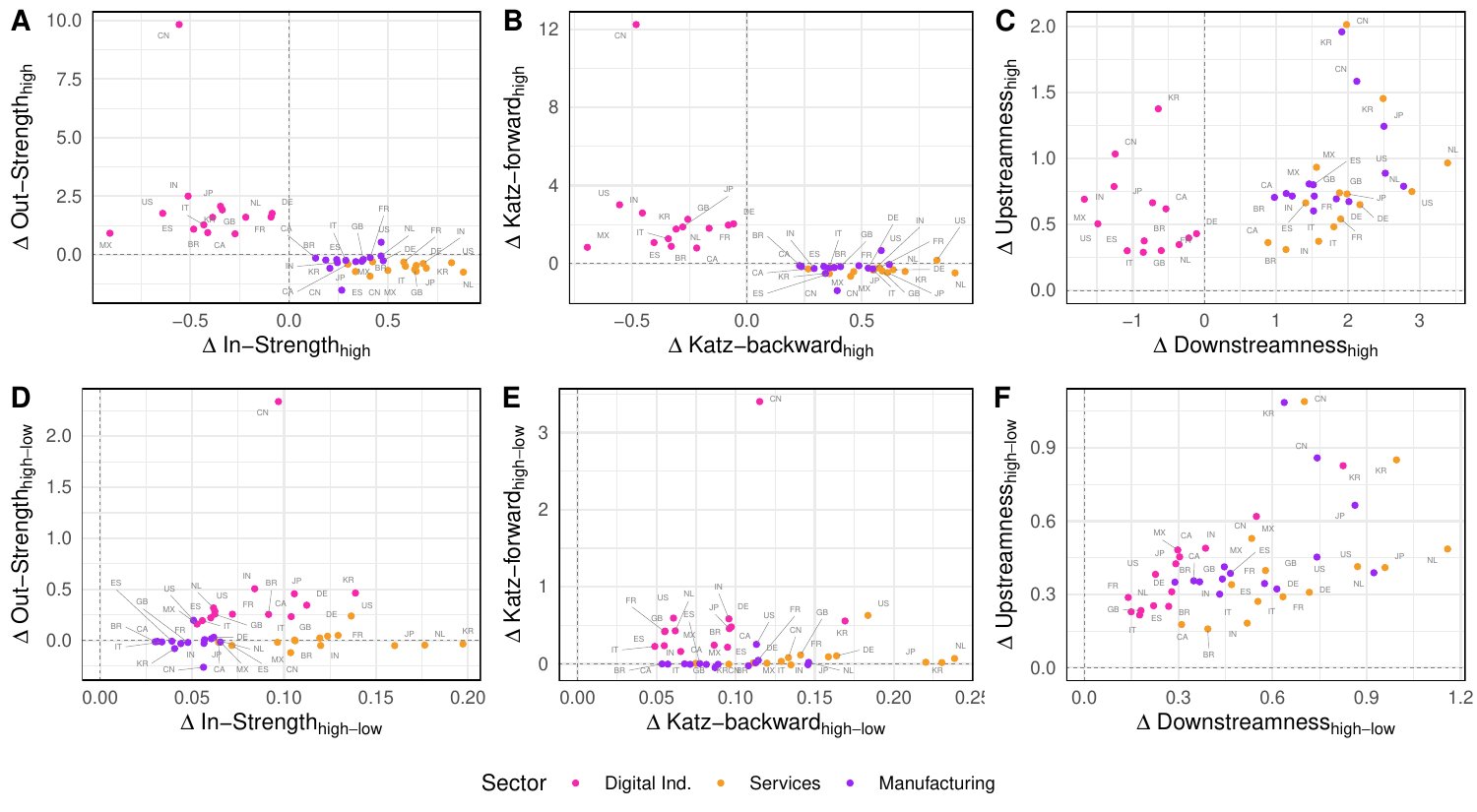}
\caption{Changes in production-network statistics for digital industries, services, and manufacturing. Upper panels (A--C): high-embeddedness scenario relative to $t_0$. Lower panels (D--F): high-embeddedness relative to low-embeddedness at the end of the simulations. Horizontal and vertical axes respectively pair in-strength with out-strength (A, D), backward with forward Katz centrality (B, E), and downstreamness with upstreamness (C, F). Network definitions are provided in Section~\ref{SM-sec:S_network_measures}.}
\label{fig:d_net_delta}
\end{figure}

Network statistics are computed at the firm level and subsequently summed within each country's economic sector. Figure~\ref{fig:d_net_delta} presents changes in network statistics for digital industries, services, and manufacturing. For any reported statistic $\xi$, the upper panels show $\Delta \xi_{high}=\xi_{high}(T)-\xi(t_0)$, whereas the lower panels show $\Delta \xi_{high-low}=\xi_{high}(T)-\xi_{low}(T)$. The latter compares the final networks under the high-embeddedness and low-embeddedness scenarios, starting from the same calibrated economy and receiving the same digital-supplier TFP shock. It captures the model response to the alternative nesting of digital inputs, including all resulting price, quantity, and labour adjustments; it is not a decomposition into separate causal channels.

Relative to the initial benchmark, digital industries show the clearest asymmetry: in-strength and backward Katz centrality decline, while out-strength and forward Katz centrality rise (Figure~\ref{fig:d_net_delta}, A--B). China stands out for the largest increases in both forward measures. The digital group thus combines lower direct and indirect input-requirement statistics with a greater weight as a supplier in the coefficient network. This aligns with common digital productivity improvement and endogenous input substitution, but the indicators do not separate their contributions. Greater forward centrality also does not imply a proportional increase in intermediate sales: technical coefficients describe input intensity, whereas sales also depend on customers' output levels.

Services and manufacturing generally display stronger backward linkages and weaker forward measures, although manufacturing includes positive forward responses in some countries (Figure~\ref{fig:d_net_delta}, A--B). Thus, output expansion can coexist with stronger intermediate-input dependence, consistent with the smaller GVA responses, without implying that every sector becomes a more important supplier.

The comparison between high and low embeddedness provides an important qualification to the digital backward response (Figure~\ref{fig:d_net_delta}, D--E). Under high embeddedness, digital in-strength and backward Katz centrality are higher, even though both remain below their initial values. These signs are compatible: high embeddedness attenuates the decline relative to $t_0$ rather than reversing it. Digital forward measures are also generally higher under high embeddedness, with China's digital industries again showing the largest contrast. The alternative nests therefore reinforce the digital supplier role while moderating the decline in its backward-linkage measures, under an identical exogenous TFP path.

Outside digital industries, services and manufacturing show much smaller differences between scenarios, with forward measures clustering closer to zero and including both positive and negative responses (Figure~\ref{fig:d_net_delta}, D--E).

The positional indicators also depend on the comparison. Relative to the initial benchmark, digital upstreamness rises while reported downstreamness falls; services and manufacturing generally rise in both measures (Figure~\ref{fig:d_net_delta}, C). Both indicators are higher under high embeddedness than under low embeddedness for these three groups (Figure~\ref{fig:d_net_delta}, F). These patterns suggest that digital industries become connected to longer production chains towards final demand while relying on shorter input paths towards their suppliers, whereas other sectors become embedded in longer production chains in both directions.

Although industries related to agriculture and food, energy, and mining are generally less exposed to digital technologies according to the taxonomies used in this study, they also experience changes in their production-network linkages. Relative to $t_0$, these sectors generally exhibit increases in in-strength and backward Katz centrality, while forward measures show more heterogeneous responses across countries, particularly in energy and mining. Differences between embeddedness scenarios are generally smaller but remain visible in both backward linkages and network-position indicators. These results suggest that digital productivity improvements propagate beyond directly exposed industries through existing production interdependencies. The corresponding results are reported in Section~\ref{SM-sec:network_other_sectors}.

\subsection{Labour compensation and reallocation}
Changes in input requirements and sectoral activity also affect labour remuneration and allocation. Comparing nominal compensation, allocated hours, and cumulative flows distinguishes the remuneration response from the reallocation that accompanies production-network recomposition.

\begin{figure}[h!]
\centering
\includegraphics[width=\linewidth]{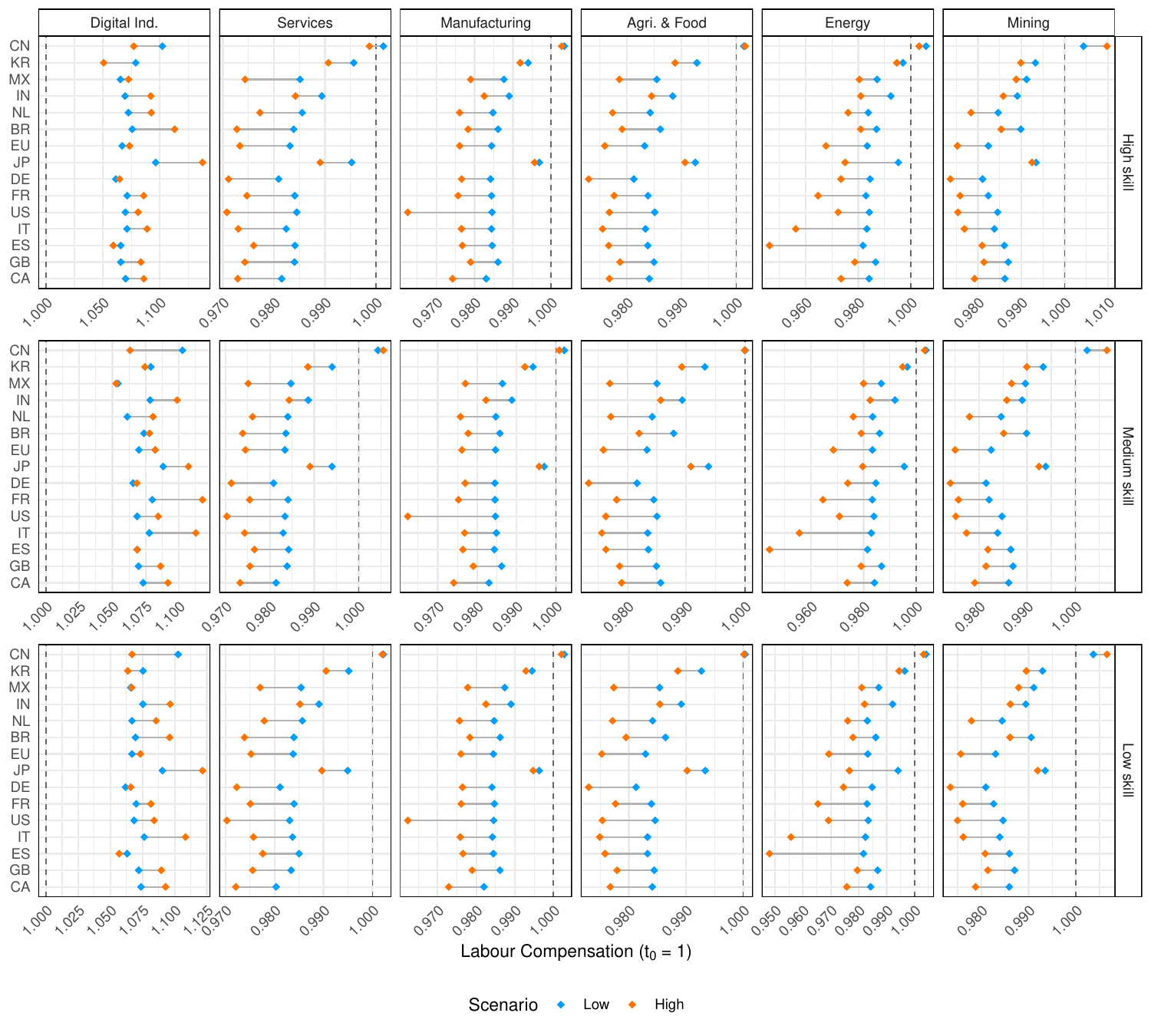}
\caption{Nominal labour compensation by country, broad economic sector, and skill group at the end of the simulation, relative to the initial benchmark ($t_0=1$). Columns show digital industries, services, manufacturing, agriculture and food, energy, and mining; rows show high-, medium-, and low-skill labour, from top to bottom. Colours distinguish the low- and high-embeddedness scenarios.}
\label{fig:compensations}
\end{figure}

Figure~\ref{fig:compensations} reports nominal labour compensation, combining wages and working hours. Digital industries exhibit increases of approximately 5--14\% across the country--skill groups displayed. The high-embeddedness scenario generally reinforces these gains, although some economies, notably China, Korea, and Spain, show lower digital compensation under high embeddedness than under low embeddedness. In contrast, nominal compensation declines in most non-digital sectors. In services and manufacturing, these reductions are often larger under the high-embeddedness scenario, despite the increases in gross output documented earlier. This divergence indicates that production expansion does not necessarily translate into higher labour remuneration within the same sector.

A decline in nominal compensation alone cannot establish a loss of working hours or real purchasing power, which also depends on consumer prices. Conversely, rising aggregate household utility does not imply gains for each household or sectoral labour group.

\begin{figure}[h!]
\centering
\includegraphics[width=\linewidth]{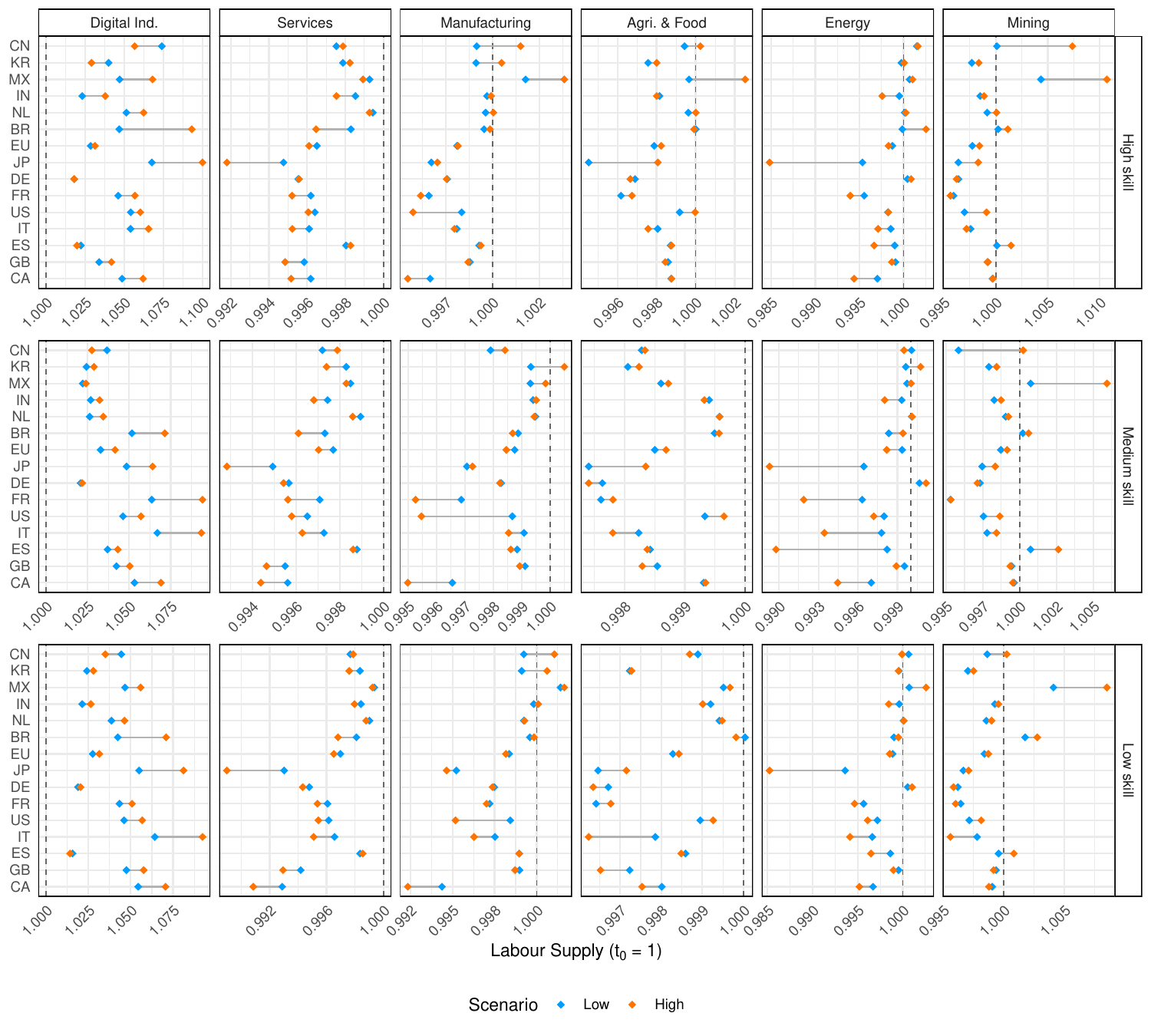}
\caption{Allocated working hours (labour supply) by country, broad sector, and skill group at the end of the simulation, relative to the initial benchmark ($t_0=1$). National skill-specific labour endowments are fixed. Columns show digital industries, services, manufacturing, agriculture and food, energy, and mining; rows show high-, medium-, and low-skill labour. Colours distinguish the low- and high-embeddedness scenarios.}
\label{fig:l_supply}
\end{figure}

Figure~\ref{fig:l_supply} shows reallocation towards digital industries across the countries and skill groups considered. Japan, Brazil, Italy, and France include increases of approximately 8--10\% in allocated digital working hours for some skill groups under the high-embeddedness scenario. Services generally lose less than 1\% of their initial hours. Manufacturing adjustments are smaller and more heterogeneous, including inflows for some skill groups in Mexico, China, and Korea. National labour endowments are fixed within each skill group, so these changes redistribute available hours across domestic industries. A small proportional reduction in a large sector such as services can accommodate a much larger percentage increase in the smaller digital sector.

Digital productivity improvements are thus accompanied by increases in both allocated hours and nominal compensation in digital industries. Elsewhere, compensation often falls proportionately more than hours, indicating a decline in nominal compensation per allocated hour. This ratio combines changes in industry wages and the allocation of hours within each broad sector; it does not isolate a pure wage effect at unchanged composition. The transformation reshapes both labour allocation and the distribution of nominal labour income.

\begin{figure}[h!]
\centering
\begin{tabular}[b]{@{}c@{}}
\textsc{\small Germany}\\[-1pt]
\includegraphics[width=0.24\linewidth]{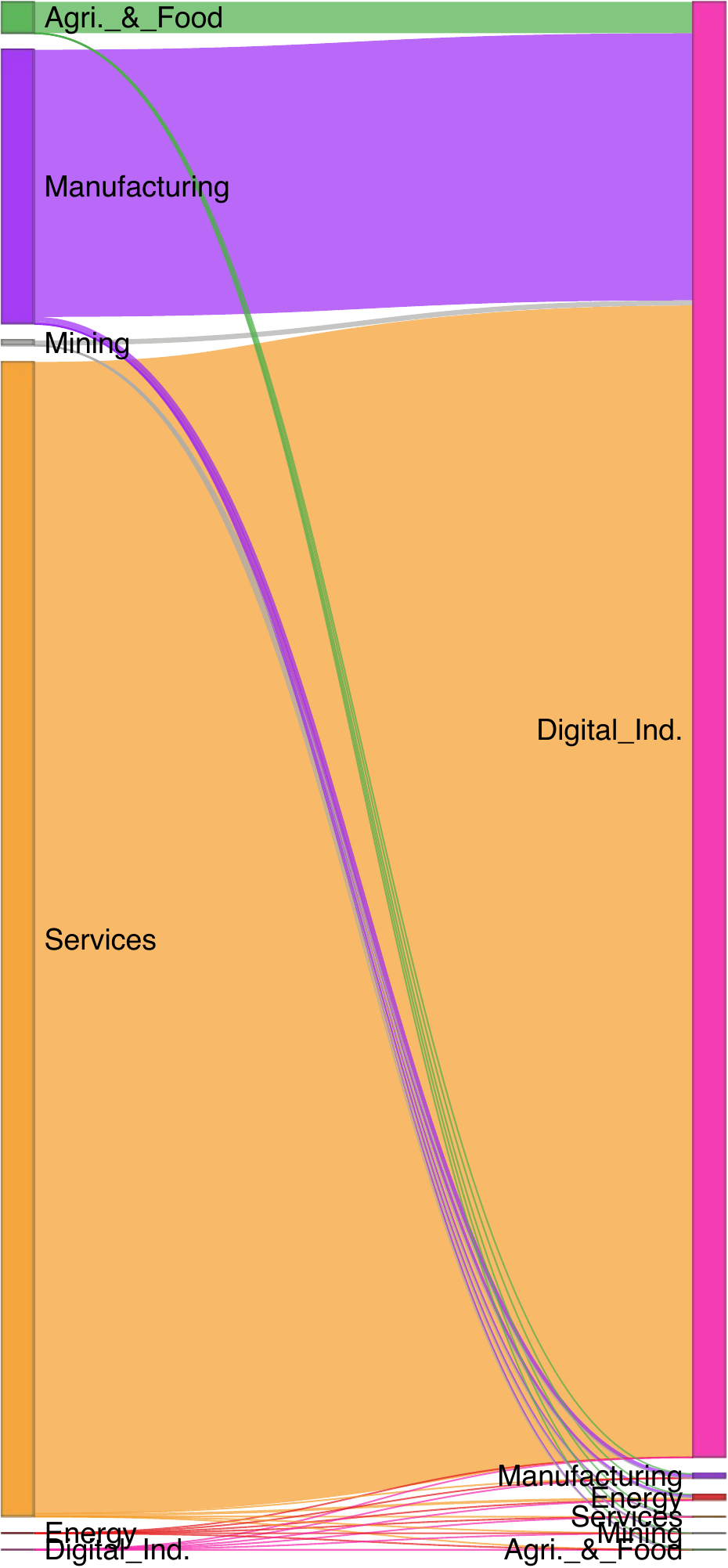}
\end{tabular}
\hfill
\begin{tabular}[b]{@{}c@{}}
\textsc{\small United States}\\[-1pt]
\includegraphics[width=0.24\linewidth]{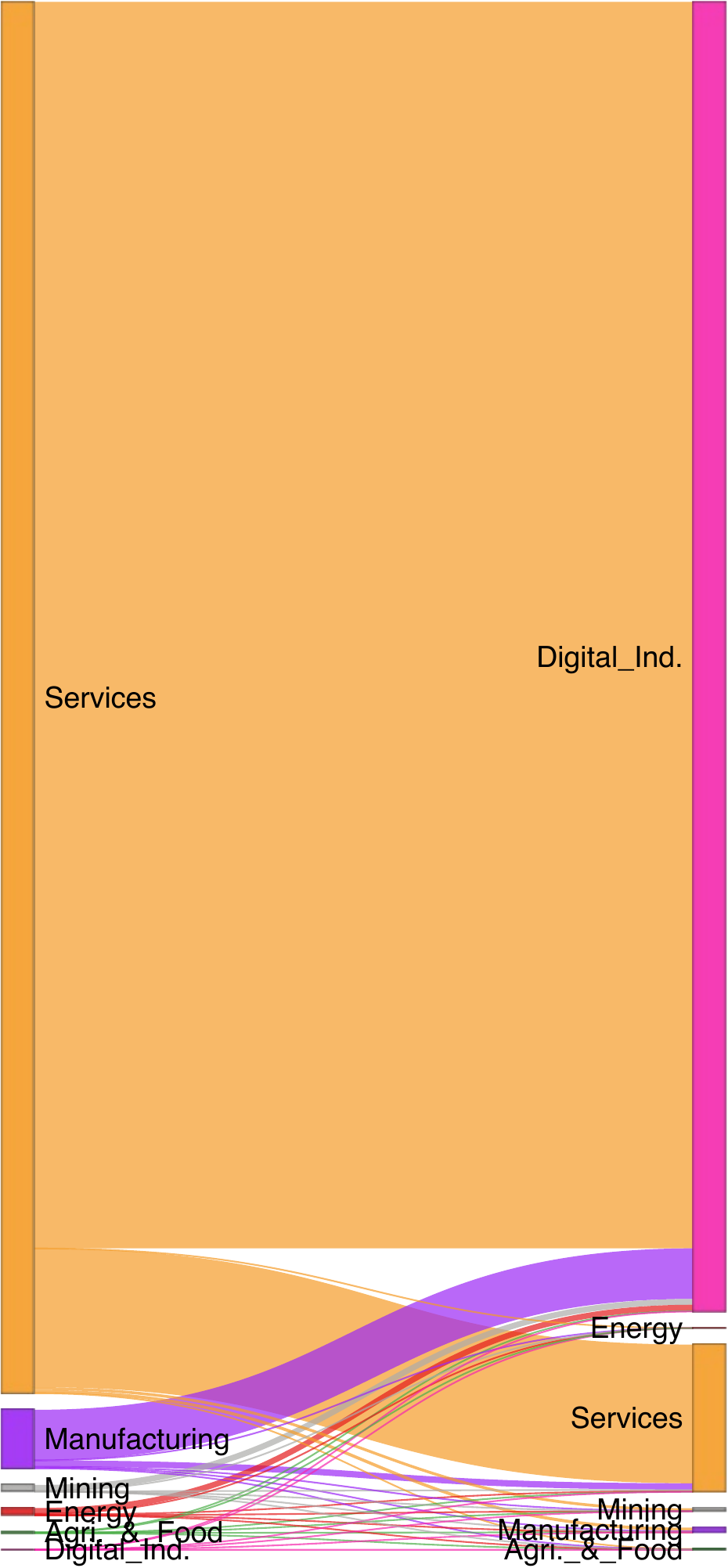}
\end{tabular}
\hfill
\begin{tabular}[b]{@{}c@{}}
\textsc{\small Korea}\\[-1pt]
\includegraphics[width=0.24\linewidth]{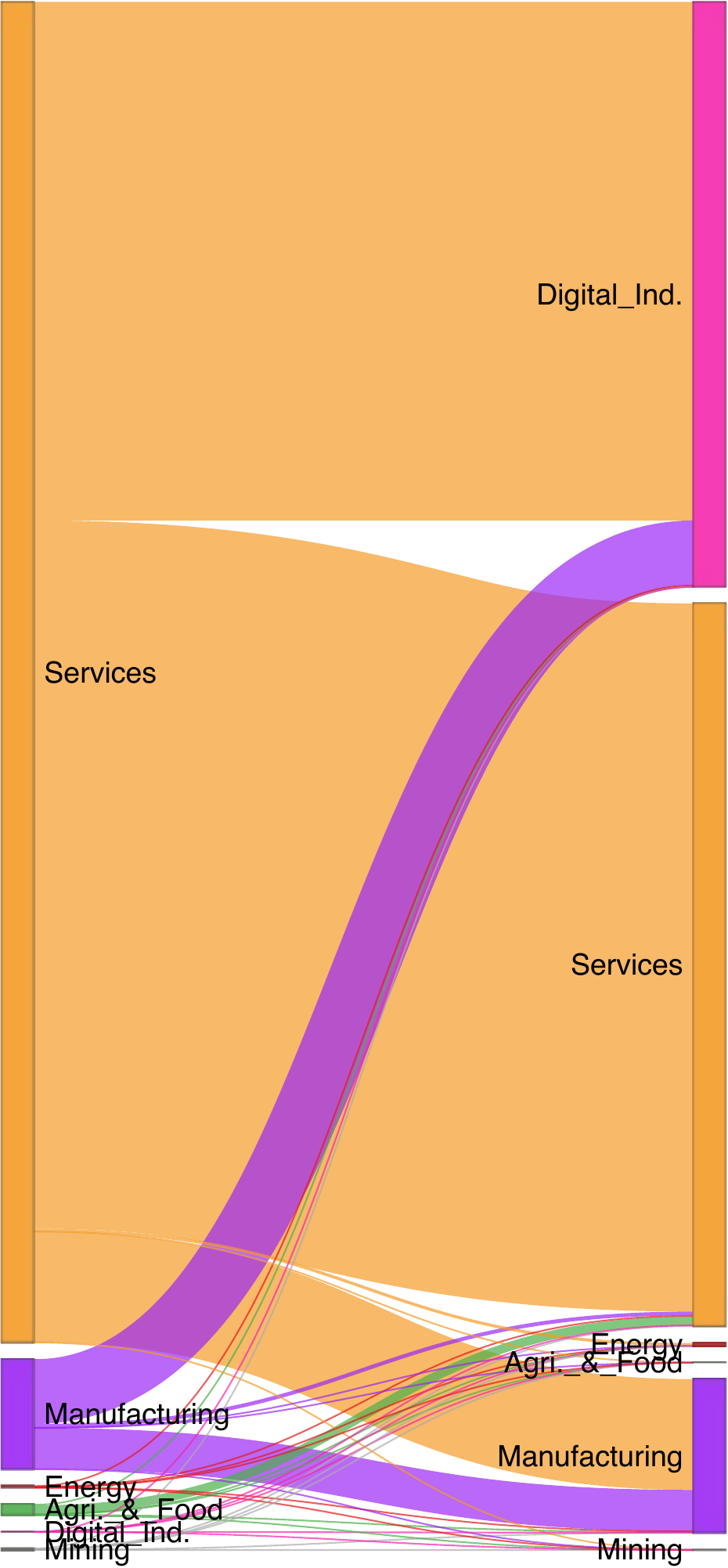}
\end{tabular}
\hfill
\begin{tabular}[b]{@{}c@{}}
\textsc{\small China}\\[-1pt]
\includegraphics[width=0.24\linewidth]{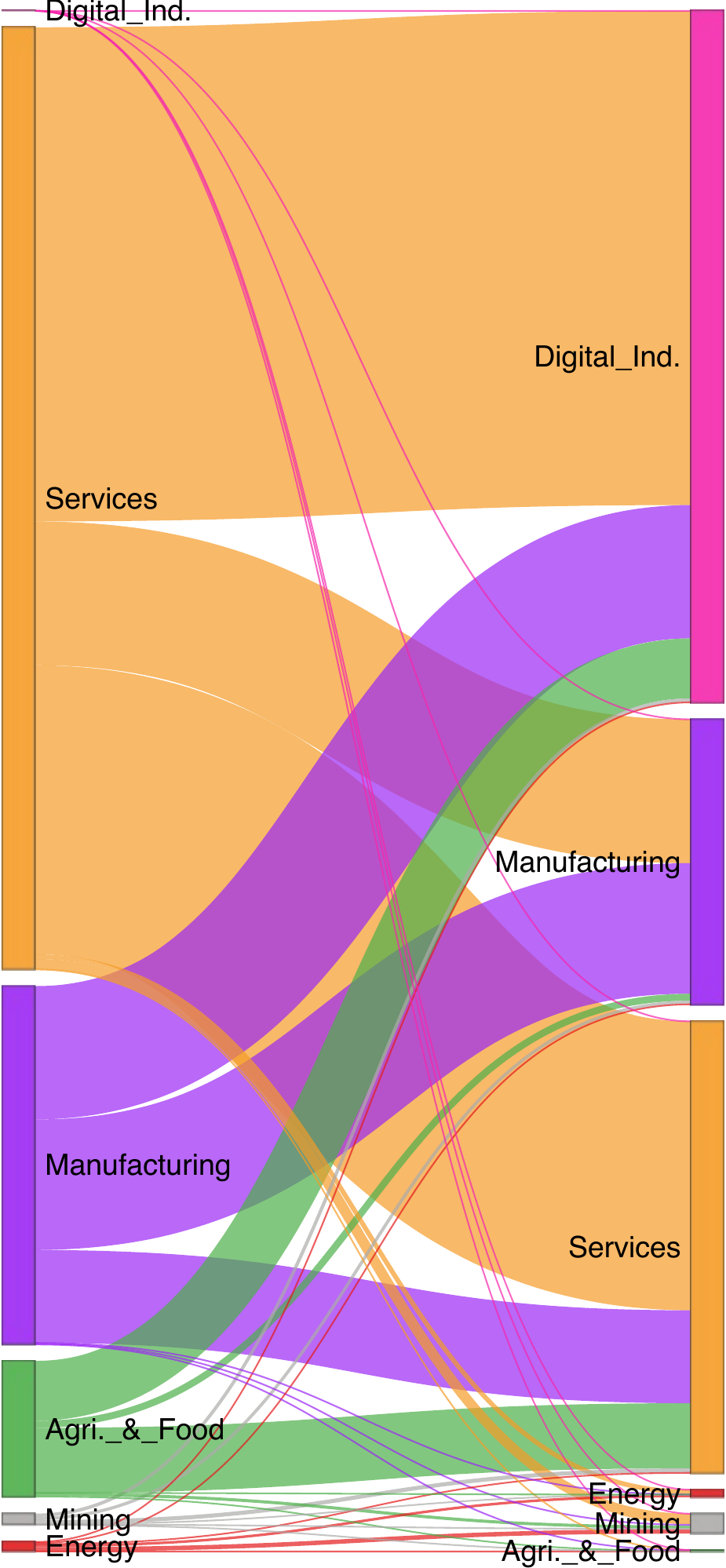}
\end{tabular}
\caption{Cumulative gross sectoral labour reallocation under the high-embeddedness scenario through the end of the simulation, for Germany, the United States, Korea, and China. Links run from origin sectors on the left to destination sectors on the right; widths represent reallocated working hours within each country. Moves between industries in the same broad sector can appear as same-sector links.}
\label{fig:sankey_labour}
\end{figure}

Figure~\ref{fig:sankey_labour} provides a more detailed picture of the sectoral reallocation process by tracing cumulative gross working-hour flows under the high-embeddedness scenario. Across the four economies considered, digital industries emerge as an important destination of labour reallocation, although the sectoral origins and composition of these flows differ substantially. In Germany and the United States, reallocation is particularly concentrated along the services-to-digital corridor, indicating that services constitute a major source of labour for the expanding digital sector. Korea exhibits a similar pattern, with services playing a prominent role, although manufacturing also contributes substantially to digital labour inflows, as in Germany. China, by contrast, shows a more diversified structure, with substantial flows from manufacturing, agriculture, and food, alongside services, and a broader distribution of destinations across digital and non-digital activities.

These country differences are consistent with the role of productive structure in shaping available reallocation routes. Mobility depends on wage differentials and the similarity of industries' input structures, so expanding digital activities need not draw workers from the same sectors in every economy. The Sankey diagrams describe the cumulative adjustment process, including movements within broad sectors; they do not measure net sectoral changes or demonstrate that remaining adjustment pressures have disappeared.

The joint expansion of digital production, compensation, and hours relates to the distinction between labour-displacing and labour-demand-enhancing effects of technological change discussed by \citet{AcemogluRestrepo2019}. Our model does not explicitly capture the automation or creation of tasks. Labour reallocation instead emerges from sectoral productivity improvements, input substitution, and adjustments transmitted through production networks.

The direction of sectoral adjustment also differs from the evidence on industrial robots presented by \citet{DauthEtAl2021}. In Germany, they find that robot-related employment losses in manufacturing are offset by employment growth in services. In our simulations, by contrast, services constitute an important source of labour moving towards expanding digital industries. Their findings concern sectoral employment responses to automation, whereas our Sankey diagrams describe cumulative gross labour flows. Nevertheless, the contrast illustrates how the sectoral patterns of labour adjustment depend on the nature of technological change and how it propagates through the production structure.

\subsection{Labour gaps and mobility frictions}
Substantial cumulative labour flows can coexist with remaining reallocation pressures. To assess these pressures, we use the labour-gap indicator defined in Eq.~\eqref{eq:labour_gap}, which expresses differences in hourly remuneration across industries in terms of working hours. Positive gaps identify workers in relatively low-wage industries who could face unemployment risks if employment in their industries contracts, and they cannot move to expanding activities. Conversely, negative gaps identify industries offering relatively higher wages and maintaining incentives to attract additional workers. These gaps therefore indicate potential labour-market adjustment pressures, rather than measuring actual unemployment or unfilled vacancies.

Note that positive and negative industry gaps offset within each country--skill group. We first measure the overall magnitude of reallocation pressures: the aggregate indicator sums positive industry gaps and expresses them as a percentage of initial working hours within the same country--skill group. 

\begin{figure}[h!]
\centering
\includegraphics[width=0.99\linewidth]{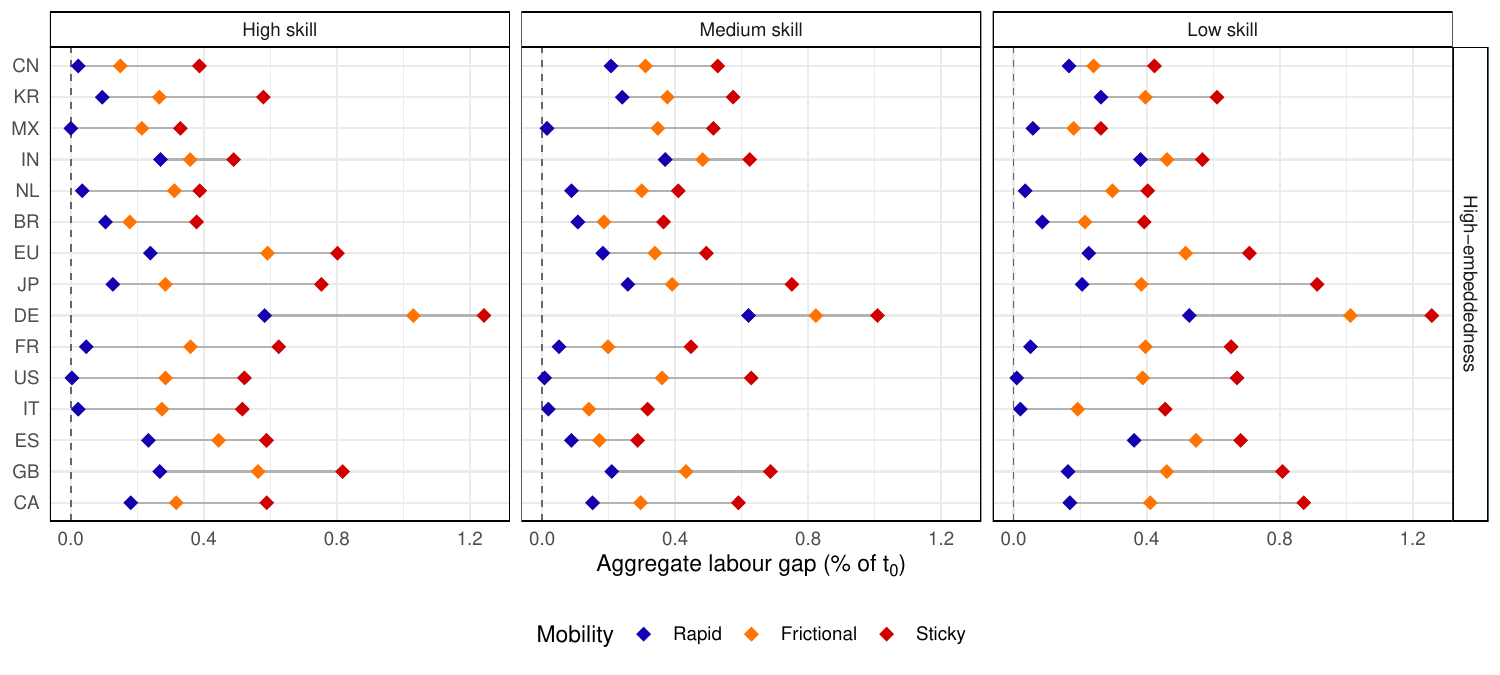}
\caption{Aggregate labour-gap diagnostic under high embeddedness at the end of the simulation, relative to initial employment ($t_0$). For each country and skill group, the sum of positive industry gaps is divided by initial employment in that same country--skill group. Colours distinguish rapid ($\tau_h=0.99$), frictional ($\tau_h=0.015$), and sticky ($\tau_h=0$) mobility. Values are percentages of initial employment, rather than percentage changes in unemployment.}
\label{fig:unemp_mob}
\end{figure}

Figure~\ref{fig:unemp_mob} compares sticky ($\tau_h=0$), frictional ($\tau_h=0.015$), and rapid ($\tau_h=0.99$) mobility under the high-embeddedness scenario. As expected, greater mobility reduces the aggregate gaps across the country--skill groups displayed, with substantial differences in the residual magnitude. Germany exhibits the largest gaps, approximately 1.0--1.3\% of initial employment under sticky mobility and roughly 0.5--0.6\% under rapid mobility. In the United States, gaps of approximately 0.5--0.7\% under sticky mobility fall close to zero under rapid adjustment. India retains noticeable gaps under rapid mobility, as does China for medium- and low-skill workers.

\begin{figure}[h!]
\centering 
\includegraphics[width=0.98\linewidth]{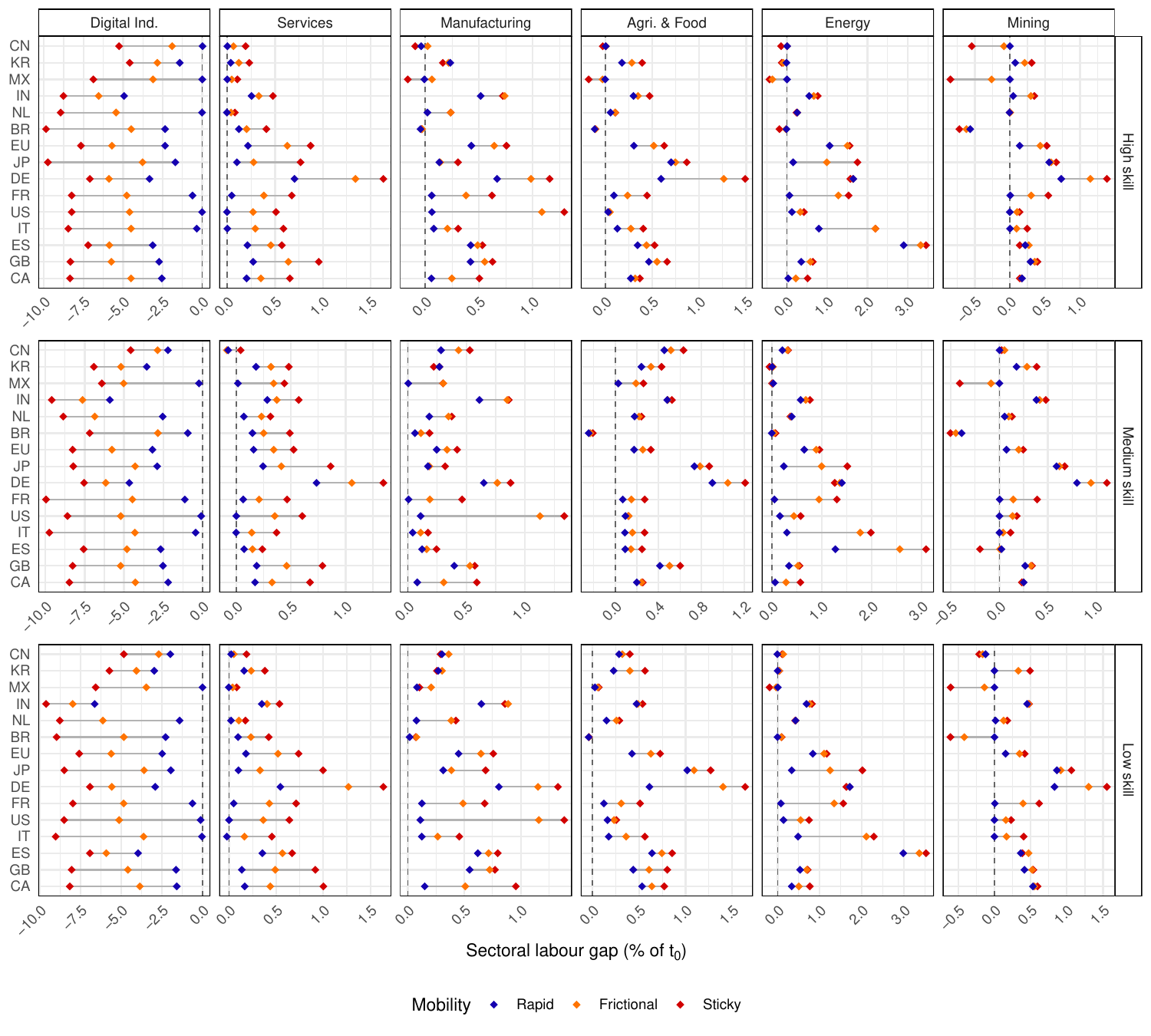}
\caption{Sectoral labour-gap diagnostic under high embeddedness at the end of the simulation. Signed industry gaps are summed within each broad sector and expressed as a percentage of initial employment in the corresponding country--sector--skill group. Columns show digital industries, services, manufacturing, agriculture and food, energy, and mining; rows show high-, medium-, and low-skill labour. Positive values indicate net hours above the equal-remuneration reference allocation, and negative values indicate hours below it; they do not measure actual vacancies or unemployment. Colours distinguish rapid, frictional, and sticky mobility.}
\label{fig:unemp_mob_sec}
\end{figure}

Figure~\ref{fig:unemp_mob_sec} shows economic sector gaps, summed with the signs of their constituent industries and normalised by initial working hours in the corresponding sector and skill group. This lets us measure net sectoral reallocation pressure while identifying which sectors have hours above or below their reference allocation.\footnote{Positive and negative disaggregated industry gaps can offset within a broad economic sector, so these figures describe its net position relative to the equal-remuneration reference.}

Digital-sector gaps are negative or close to zero. Under sticky mobility, their absolute magnitudes are often around 8--10\% of initial sectoral employment, with some exceeding 10\%; faster mobility substantially reduces them. The frictional benchmark therefore combines inflows into digital industries with continuing inward reallocation pressures relative to the reference allocation. 

Services generally exhibit positive gaps, particularly in Germany, identifying a potential source of hours for reallocation relative to the same reference. Manufacturing outcomes are more heterogeneous. In the United States, the positive gap under sticky mobility is approximately 1.3--1.4\% of initial manufacturing employment across skill groups, but falls considerably under rapid mobility. Aggregate expansion can therefore coexist with opposing reallocation pressures across sectors.

Agriculture and food, energy, and mining show both signs, depending on the country and skill group. Energy in Spain stands out, with positive gaps of approximately 3--3.5\% of initial sectoral employment under sticky mobility. Mining shows positive gaps in Germany and Japan and negative gaps in Brazil and some skill groups in Mexico and China. Faster mobility generally reduces the magnitude of labour gaps. However, net gaps at the broad-sector level need not decline monotonically with mobility, because positive and negative gaps across constituent industries can offset one another.

Aggregate gains from digital productivity growth coexist with an uneven adjustment of labour across industries. Digital activities attract working hours, yet these inflows do not eliminate the pressures for further reallocation. Gaps relative to the national average wage reference persist in both digital and non-digital activities. Faster mobility reduces their aggregate magnitude, but the remaining pressures differ across countries and skill groups. The labour implications of digital expansion therefore depend on both which activities grow and how readily labour can move between them over the simulated horizon.

\subsection{Sensitivity analysis}
Having examined the results under the benchmark parameterisation, we now assess whether the main world-level aggregate patterns and differences between technological scenarios persist under alternative parameter values. This sensitivity analysis also addresses uncertainty regarding the magnitude of digital productivity improvements and the substitution possibilities between digital inputs and primary factors.

We explore a two-dimensional parameter grid combining cumulative digital-supplier TFP increases, $g_D \in \{10\%, ..., 40\%\}$, with value-added substitution elasticities, $\sigma_{\mathrm{VA}} \in \{0.50, ..., 1.25\}$. All other parameters, including frictional labour mobility and the shock duration, remain fixed. We present two complementary sensitivity exercises comparing the high-embeddedness scenario against the initial benchmark economy and the low-embeddedness one. 

\begin{figure}[h!]
\centering
\includegraphics[width=0.99\linewidth]{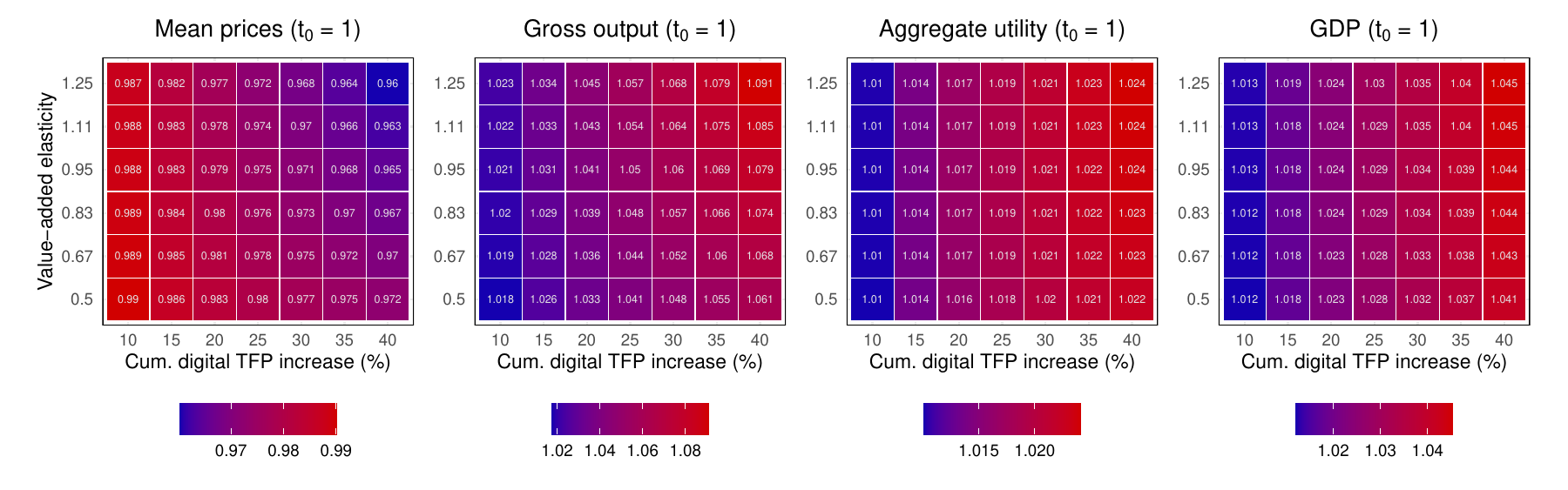}
\includegraphics[width=0.99\linewidth]{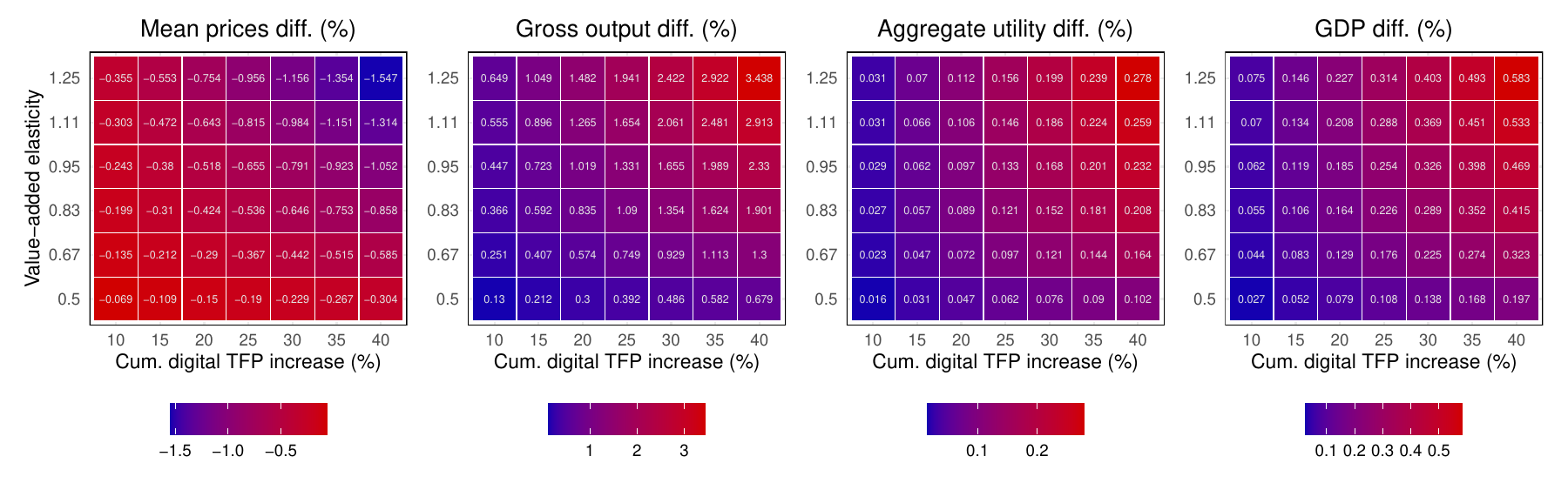}
\caption{Sensitivity of global aggregate outcomes to cumulative digital-supplier TFP improvements and value-added substitution elasticities at the end of the simulation ($T$). Upper panel: Mean prices, gross output, aggregate household utility, and GDP under the high-embeddedness scenario, expressed as ratios to their initial benchmark values ($t_0=1$). Lower panel: Differences between the high- and low-embeddedness scenarios, expressed as percentages of the low-embeddedness outcomes. Columns within each grid vary cumulative digital TFP growth, and rows vary the value-added substitution elasticity. All other parameters remain fixed.}
\label{fig:tfp_va_grid}
\end{figure}

The upper panel of Figure~\ref{fig:tfp_va_grid} reports global aggregate outcomes under the high-embeddedness production structure, relative to the initial benchmark ($t_0=1$). Across the parameter grid, mean prices remain below their initial level, while gross output, GDP, and aggregate household utility remain above it. Larger digital productivity improvements systematically strengthen these responses at each tested elasticity. For example, at a value-added elasticity of 1.25, increasing the cumulative TFP shock from 10\% to 40\% raises the gross-output index from approximately 1.023 to 1.091 and the GDP index from 1.013 to 1.045. The corresponding mean-price index falls from about 0.987 to 0.960. The magnitude of the response, however, is not equally sensitive to the substitution elasticity across variables. At a 40\% TFP increase, raising the value-added elasticity from 0.50 to 1.25 increases the gross-output index from approximately 1.061 to 1.091 and lowers the mean-price index from 0.972 to 0.960. By comparison, the GDP index changes only from about 1.041 to 1.045, and aggregate household utility changes from 1.022 to 1.024. Thus, greater substitution possibilities amplify gross production and price adjustment much more than they amplify aggregate value added and utility.

The lower panel of Figure~\ref{fig:tfp_va_grid} compares the high- and low-embeddedness scenarios, expressed as percentages of the low-embeddedness outcomes. High embeddedness produces higher gross output, GDP, and aggregate utility and lower mean prices throughout the displayed grid. The differences generally grow in magnitude with both parameters, particularly for output and prices. At a 40\% TFP increase and an elasticity of 1.25, gross output is approximately 3.44\% higher under high embeddedness, GDP is 0.58\% higher, and aggregate household utility is 0.28\% higher; mean prices are approximately 1.55\% lower. Alternative nesting therefore affects gross production and prices more strongly than aggregate value added and utility.

Overall, these results show that the direction of the responses remains stable, while their magnitude depends on the assumed technological conditions, with the high-embeddedness production structure having more pronounced effects on gross output and prices than on GDP and household utility. This provides evidence that the findings discussed before are not limited to the selected benchmark parametrisation.\footnote{Although results are presented in aggregate form, the same qualitative trends also hold when the data are disaggregated at the country level; these results are available upon request.}

\section{Concluding remarks}  
\label{sec:concluding_remarks}

This paper examines how productivity improvements originating in digital industries propagate through production networks and reshape the sectoral allocation of labour, using an agent-based model simulating economic dynamics in global production networks. Leveraging nested CES specifications of industrial production, we consider two alternative technological structures: a low-embeddedness scenario, in which digital inputs remain in a separate intermediate-input nest, and a high-embeddedness scenario, in which these inputs enter an augmented capital--labour nest in potentially exposed industries.

We find that digital productivity gains generate uneven economic expansion across countries and industries. In both technological structures, as digital productivity rises, average prices fall while gross output, GDP, and household utility increase. At the sectoral level, responses are particularly heterogeneous. Under high embeddedness, digital gross output roughly doubles in Korea and China, while non-digital activities typically experience much smaller gains and, in some countries-sectors, contractions in value added. Gross output also tends to increase more than GDP at the aggregate level and more than value added at the sectoral level, indicating that greater production activity along interconnected supply chains does not necessarily translate into comparable gains in value added. Moreover, these differences are accompanied by uneven changes in labour remuneration across industries.

These aggregate patterns are qualitatively robust to alternative assumptions about the magnitude of digital productivity improvements and the value-added substitution elasticity. Although these parameters affect the magnitude of the responses, the direction of the main aggregate effects remains unchanged across the ranges considered.

Beyond expanding production, digital productivity improvements also reshape interindustry production linkages. Relative to the initial benchmark, digital industries generally reduce their intermediate-input requirements per unit of output while becoming more important direct and indirect suppliers under both embeddedness scenarios. Their position within production networks also changes: reported upstreamness increases while downstreamness decreases, whereas services and manufacturing generally exhibit increases in both measures. These changes accompany a growing role for digital industries as intermediate-input suppliers, alongside a declining intensity of their own intermediate-input requirements. The magnitude of these changes, however, depends on the scenario: compared with low embeddedness, high embeddedness attenuates the decline in digital industries' backward-linkage measures and generally amplifies the increase in their forward-linkage measures. Thus, the same productivity shock can generate different degrees of production-network recomposition depending on how digital inputs are combined with capital and labour.

A redistribution of labour accompanies changes in production networks. Digital industries attract working hours and register higher nominal labour compensation, while other activities, especially services, provide much of the reallocated labour in several economies. The origins of these flows differ across countries, reflecting their productive structures and the model's wage and technological-similarity constraints. However, inflows into expanding industries do not eliminate adjustment pressures. Digital industries continue to offer relatively high hourly wages, indicating persistent incentives to attract additional workers, while other activities often exhibit incentives for workers to move elsewhere. Under sticky mobility, the corresponding labour gaps in digital industries are often equivalent to 8--10\% of their initial working hours. Faster mobility reduces these wage-driven reallocation pressures but does not necessarily eliminate them, with residual imbalances varying across countries and skill groups.

In short, our results suggest that the economic consequences of digital transformation cannot be assessed from aggregate productivity or GDP alone. They also depend on the structure of intermediate production and on the capacity of labour to move between activities. Our analysis is deliberately conditional: productivity improvements are exogenous, the feasible supplier network is fixed, national skill-specific labour endowments do not grow, and neither technology adoption nor task-level automation is modelled explicitly. Within these boundaries, the model identifies how alternative technological structures can generate different patterns of production-network recomposition and persistent labour-adjustment pressures. Digital transformation is therefore not only a question of how productivity gains spread through the economy, but also of whether workers can adapt to the changing structure of production, which highlights the importance of policies that facilitate labour mobility and reduce barriers to sectoral adjustment.

\section*{Acknowledgements}
The authors gratefully acknowledge funding from the European Union's Horizon Europe research and innovation programme under grant agreement No.~101132559 (ST4TE--Strategies for Just and Equitable Transitions in Europe). The funding source had no role in the design of the study, the analysis or interpretation of the results, the preparation of the manuscript, or the decision to submit it for publication.

\clearpage
\begin{center}
\section*{\small Supplementary Material }
\vspace{10pt}
{\LARGE Structural economic change under the digital transformation}
\end{center}
\vspace{10pt}

\setcounter{section}{0}
\setcounter{table}{0}
\setcounter{figure}{0}

\renewcommand{\thesection}{SM.\arabic{section}}
\renewcommand{\thetable}{SM.\arabic{table}}
\renewcommand{\thefigure}{SM.\arabic{figure}}
\renewcommand{\theHsection}{SM.\arabic{section}}
\renewcommand{\theHtable}{SM.\arabic{table}}
\renewcommand{\theHfigure}{SM.\arabic{figure}}
\newcolumntype{P}[1]{>{\raggedright\arraybackslash}p{#1}}

\section{Economic model details}
\label{sec:model_SM}\label{SM-sec:model_SM}

We use an agent-based implementation to study disequilibrium responses to digital-sector productivity improvements based on \cite{gualdi2016emergence, mandel2021monetary, duenas2026labour}. 

\subsection{General equilibrium framework}

The set of goods $\M$ in the economy corresponds to the output of the different types of agents: the three types of skilled labour, the capital good, and the output of each type of industry, all differentiated by country. Thus, there are $M=|\M|=C(L+4)=8183$ such goods and therefore the same number of agents. We shall denote by $\F \subset \M$ the set of industries (and the corresponding outputs) and by $\H=\H_L\cup \H_K$ the set of households, comprising workers and capitalists. 

Each firm $i$ produces its differentiated good using capital provided by the domestic capitalist, the different types of labour provided by domestic workers, and a combination of domestic and international inputs. More precisely, the production technology of each firm $i$ is described by a nested constant elasticity of substitution (CES) production function of the form:
\begin{equation} \label{prodfun}
f_{i}(x_i) = \Lambda_i \phi_i(x_i)
\end{equation}
with, 
\begin{equation}
\phi_i(x_i) =
\left[
  \sum_{s \in S_i}
  \left(
    \sum_{j \in s} \beta_{ji} x_{ji}^{\rho_s}
  \right)^{\theta / \rho_s}
\right]^{1/\theta}.
\end{equation}
where $\phi_i$ denotes physical production technology, $\Lambda_i$ total factor productivity (TFP), $x_i=(x_{ji})_{j\in\M}$ the input vector, and $S_i$ the partition of inputs into nests. At the calibration benchmark, $\Lambda_i=1$ for every firm. The standard production partition comprises value added (domestic capital and the three skill-specific labour inputs) and seven intermediate-input nests: agriculture and forestry, energy, manufacturing, digital inputs, services, fossil inputs, and mining. Under high embeddedness, digital inputs enter the augmented value-added nest only in exposed industries, as specified in the main text. Input classifications are reported in Table~\ref{tab:S_industries}.

Each type of worker $h \in \H_L$ provides a fixed amount of work $e_h \in \R_+$ and each capitalist $h \in \H_K$ provides a fixed amount of capital services $e_{h} \in \R_+$ every period. Households also have nested CES preferences of the form
\begin{equation} \label{utfun}
u_{h}(x_h) = 
\left[
  \sum_{s \in S'}
  \left(
    \sum_{j \in s} \beta'_{jh} x_{jh}^{\rho'_s}
  \right)^{\theta' / \rho'_s}
\right]^{1/\theta'}.
\end{equation}
where $x_h$ is household $h$'s consumption vector and $S'$ is the common household partition. Its eight nests are agriculture and food, energy, manufacturing, digital inputs, transport, services, waste, and mining. Household nests are unchanged across modelling scenarios and are distinct from production nests.

We denote the economic system introduced above as $\E(\M,\beta,\rho,\theta).$  In this setting, a general equilibrium is standardly defined as follows.

\begin{defi} A general equilibrium of the economy $\mathcal{E}(\mathcal{M},\beta,\rho,\theta)$ is a 
collection of prices $p^* \in \R^M_+$, production levels 
$y^* \in \R^M_+$ and commodity flows $(x^*_{j,i})_{j,i\in \mathcal{M}} \in \R^{M\times M}_+$ such that:
\begin{enumerate}

\item Consumers maximise their utility under their budget constraint. That is for all $h \in \mathcal{H},$ $x^*_h \in \R^M_+$ is a solution to 
$$\left\{\begin{array}{c} \max u_h(x_h) \\  \\
\text{s.t }   p^* \cdot x_{h} \leq  p^*_h e_h \end{array}\right.$$

\item Firms maximise profits. That is for all $i \in \mathcal{F},$ 
$(y^*_i,x^*_i) \in \R_+ \times \R^M_+ $ is a solution of 
$$\left\{\begin{array}{c} \max  p^*_i y_i- p^* \cdot x_{i}  \\  \\
\text{s.t }  f_i(x_i)\geq y_i \end{array}\right.$$

\item Markets clear. That is, for every $j \in M,$ one has 
$$y^*_j = \sum_{i=1}^M x^*_{j,i}.$$
Furthermore, for all $j \in \mathcal{H},$ $y^*_j=e_j$ to account for the inelastic supply of labour and capital services in the economy.
\end{enumerate}
\label{monequi}
\end{defi}

\begin{remark}
 Note that, given constant returns in production technologies, there are no profits to be distributed at equilibrium. Namely, one has for all $j \in \mathcal{F}$
 $$p^*_j y^*_{j} =p^* \cdot x^*_{j} $$
The non-profit condition implies, in particular, that households' income is entirely determined by their supply of labour or capital services.
\end{remark}

As highlighted in the recent macroeconomic literature \cite[see, e.g.,][]{baqaee2019macroeconomic, carvalho2019production}, the general equilibrium of this economy defines nominal input--output flows between each pair of agents. These flows provide the equilibrium expenditure-share representation used to initialise the model. The structure of this network can be captured by a column-stochastic matrix $A^*=(a^*_{i,j})_{i,j \in \mathcal{M}} \in \R^{M \times M}_+$ whose coefficient $a^*_{i,j}$ captures the share of $j$'s equilibrium spending directed towards $i,$ that is  
\begin{equation} 
a^*_{i,j}=\dfrac{p^*_i x^*_{i,j}}{p^* \cdot x^*_{\cdot,j}}.
\end{equation} 
In particular if $j \in \mathcal {F}$ and $i \in \mathcal {F},$ $a^*_{i,j}$ represents the amount of intermediate consumption spent on good $i$ per unit of revenue of firm $j$ while for $i \in \mathcal {H}$, $a^*_{i,j}$ represents the value added received by household $i$ per unit of revenue of firm $j.$

\subsection*{Agent-based dynamics}

Prices, inventories, expenditure shares, and monetary balances adjust in discrete periods $t\in\mathbb N$. For each agent $i\in\M$, the state evolves dynamically through the following sequence of interactions 
\begin{enumerate}
\item  Each agent $i \in \M$ receives the nominal demand $\sum_{j \in \M} a^t_{i,j} m^t_j.$ 

\item Agents adjust their prices frictionally towards their market-clearing values according to: 
\begin{equation}
p_i^t=\tau_p \overline{p}_i^t+ 
(1-\tau_p) p^{t-1}_i \label{pricefric}\;;
\end{equation}
where $\tau_p \in[0,1]$ is a parameter measuring the speed of price adjustment and $\overline{p}_i^t$ is the market-clearing price defined as follows. Given the nominal demand $\sum_{j \in \M} a^t_{i,j} 
m^t_j$ and the output stock $q_i^t,$ the market clearing price for agent $i$ is: 
\begin{equation}\label{price}
\overline{p}_i^t=\dfrac{\sum_{j \in 
\M} a^t_{i,j} m^t_j}{\nicefrac{q_i^t}{\chi}} \;. 
\end{equation} 
where $\chi$ is the target inventory-to-output ratio.

\item Whenever $\tau_p<1,$ markets fail to clear (except if the system is at a stationary equilibrium). In case of excess demand, we assume that buyers are rationed proportionally to their demand, i.e. the quantity of good $j$ purchased by agent $i$ is given by:
\begin{equation} 
x^t_{j,i} =\min\left(1, \frac{q^t_j/\chi}{\sum_{k \in \M} a^t_{j,k}  m^t_k/p_j^t}\right) \frac{a^t_{j,i}m^t_i}{p_j^t}  \;.
\label{input} 
\end{equation}

\item Firms update their inventory by producing new output and accounting for sales:
\begin{equation}
q^{t+1}_i=q^t_i-\sum_{k \in \M} x_{i,k}^t + f_i(x_i^t) 
\label{prod}
\end{equation}

\item Agents update their monetary balances on the basis of their purchase of inputs and their sales of output\footnote{Equation~\eqref{evmoney} can be interpreted as assuming that firms have myopic expectations about their nominal demand (i.e. firms assume they will face the same nominal demand next period) and target a balanced budget (net of labour and capital costs paid to households).}, i.e.
\begin{equation}
m^{t+1}_i=m^{t}_i  + \sum_{k' \in \M} p_i^t x_{i,k'}^t - \sum_{k \in \M} p_k^t x_{k,i}^t  
\label{evmoney} 
\end{equation}

\item Agents adjust frictionally their input/consumption combinations towards the cost-minimising value: 
\begin{equation} 
a_{i}^{t+1}=\tau_{w} \overline{a}_{i}^{t} +(1-\tau_{w}) a_{i}^t \;;
\label{optweightfric}
\end{equation}
where $\tau_{w} \in[0,1]$ measures the speed of technological/preferences adjustment and $\overline{a}_{i}^{t} \in \R^M$ denotes the vector of optimal expenditure shares for agent $i$. For firms, these shares maximise production, while for households they maximise utility, given prevailing prices. Formally, the optimal expenditure shares are defined by\footnote{The fraction in Eq.~\eqref{optinput} is taken coordinate-wise.}:
\begin{equation} 
\overline{a}_{i}^{t}:= \operatorname*{\text{argmax}}_{\{a_i \mid \sum_{j\in\M} a_{j,i}=1\}} g_i\left(\frac{a_{i}}{p^t}\right) \:,
\label{optinput}
\end{equation}
where $g_i=f_i$ if $i\in\F$ and $g_i=u_i$ if $i\in\H$ for households.

\item Productivity shock: firms' productivity $\Lambda_{i,t}$ obeys the exogenous innovation path. Here, we assume TFP increases only for digital suppliers, while productivity in all other industries remains fixed. 
\end{enumerate}

The agent-based dynamics have been extensively analysed in previous contributions \cite[see, e.g.,][for details]{gualdi2016emergence,mandel2021monetary}. If the price adjustment is fast, the economy converges to the underlying general equilibrium. If price adjustment is slow, inventories carry the burden of adjustment and the economy is characterised by a synchronised state with large ``cyclical volatility''. In empirical applications \citep[see, e.g.,][]{mandel2020economic, mandel2023disequilibrium}, considering out-of-equilibrium dynamics allows us to provide a precise description of shock propagation mechanisms and of the resulting macroeconomic response.

\subsection{Labour market model}

To capture labour reallocation following the digital productivity shock, we allow wages and labour supply to vary across industries within each country and skill group. Mobility is frictional and depends on wage differentials and the transferability of skills, proxied by similarity in industry input structures at the calibration benchmark. National labour endowments remain fixed.

More formally, we represent each worker type $h \in \H_L$ as a continuous mass split across the different sectors/industries in the country. Namely, for a given country, we denote by $\ell^t_{h,i} \in \R_+$ the labour supplied by workers of type $h$ to industry $i$ in period $t,$ which is such that $\sum_{i \in \F} \ell^t_{h,i}= e_h.$ We further denote by $p^t_{h,i} \in \R_+$ the wage for workers of type $h$ in industry $i$. As in \cite{duenas2026labour}, in the baseline model we assume wages are set independently in each sector, accounting for industry-specific labour supply and demand. Specifically, we extend step 2 of the baseline model as follows. 

\begin{enumerate}[label=2'.]
\item The wage updating mechanism for industry $i$ and worker type $h$ is given by:  
\begin{equation}\label{labourpricefric}
p^t_{h,i}=\tau_p \overline{p}^t_{h,i} + (1-\tau_p) p^{t-1}_{h,i} 
\end{equation}
where $\overline{p}^t_{h,i}:= \nicefrac{a^t_{h,i} m^t_i}{\ell^t_{h,i}}$ is the clearing price on the labour market for worker-type $h$ in sector $i$.
\end{enumerate}

The second major extension we consider is that labour supply across sectors evolves dynamically. This evolution is driven by the magnitude of the wage differential between sectors and constrained by the similarity of skills/occupations across industries. More specifically, let $\widetilde p_h^t=\sum_{i\in\mathcal F_c}p_{h,i}^t\ell_{h,i}^t/e_h$ denote the hours-weighted average wage within each country and skill group. Wages below (above) this average create incentives for workers to move out of (into) an industry. To quantify these reallocation pressures, we define a reference allocation that equalises hourly remuneration across industries:
\begin{equation} 
\overline{\ell}^t_{h,i}:= \ell^t_{h,i} \dfrac{p^t_{h,i}}{\tilde{p}^t_h}
\end{equation} 
The corresponding labour gap is defined as
\begin{equation} \label{eq:labour_gap}
\Delta \ell^t_{h,i}:= (\ell^t_{h,i}-\overline{\ell}^t_{h,i}) \;. 
\end{equation}
Positive gaps indicate potential pressures for labour outflows, while negative gaps indicate incentives for additional labour inflows. These gaps do not measure observed unemployment or unfilled vacancies.

We measure the capacity of workers to transition from sector $i$ to sector $j$ through a sectoral similarity index $s_{i,j}$, defined as the cosine similarity between sectoral input vectors:
\begin{equation}
s_{i,j} = \frac{z^*_i \cdot z^*_j}{\|z^*_i\| \, \|z^*_j\|}
\end{equation}
where $z^*_i$ and $z^*_j$ denote the columns corresponding to sectors $i$ and $j$ in the baseline real input-coefficient matrix $\mathbf Z^*=(a^*_{i,j})_{i,j\in\mathcal F}$, obtained as the firm-to-firm submatrix of $A^*$. Because all prices are normalised to one in the reference equilibrium year, these coefficients can be interpreted in real terms. The numerator is the Euclidean inner product of the two vectors and $\|\cdot\|$ denotes the Euclidean norm. Hence, higher values of $s_{i,j}$ indicate greater similarity in input structures across sectors and are therefore associated with a higher likelihood of labour mobility, as workers can more easily transfer their skills across technologically related activities.

Sector similarity generates a relatively dense network of potential labour flows. To restrict potential labour flows, we retain only industry pairs with similarity above a common threshold $\bar{s}=0.2$. The resulting country-level mobility networks remain connected, allowing potential indirect transitions between industries over successive periods, conditional on wage incentives. This restriction aims to reflect the empirical evidence that inter-industry mobility is selective and concentrated among technologically related activities \citep{neffke2017inter,farinha2019drives}.

We then extend the baseline model with the following dynamics of labour supply.
\begin{enumerate}[label=8.]
\item The supply of labour by worker type $h$ in sector $i$ is given by
\begin{equation}\label{eq:labour_supply}
 \ell^{t+1}_{h,i}= \ell^{t}_{h,i} + \tau_h
 \left\{
\begin{array}{ll} 
  \sum_{\{j \mid \Delta \ell^t_{h,j} >0 \}} \Psi^t_{j,i},
  & \text{if } \Delta \ell^t_{h,i} < 0, \\[1ex]
 (-)\sum_{\{j \mid \Delta \ell^t_{h,j} <0 \}} \Psi^t_{i,j},
  & \text{if } \Delta \ell^t_{h,i} > 0 .
 \end{array}\right.
\end{equation}
where $\tau_h$ is a parameter controlling the speed of adjustment of labour supply, while the magnitude of adjustment $\Psi^t_{i,j}$ is determined as follows:

Let $i$ be an industry with a positive labour gap, so that $\Delta \ell^t_{h,i}>0$. The potential flow of workers of type $h$ from sector $i$ to sector $j$ is given by
\[
\psi^t_{i,j} = \Delta \ell^t_{h,i} 
\frac{\max(p^t_{h,j}-p^t_{h,i},0)s_{i,j}}
{\sum_k \max(p^t_{h,k}-p^t_{h,i},0)s_{i,k}},
\]
For simplicity, we removed the skill subindex $h$. We then use the expected total inflow $\sum_k \psi^t_{k,j}$ into industry $j$ to scale these flows so that they do not exceed the magnitude of the negative labour gap in destination industry $j$, where $\Delta \ell^t_{h,j}<0$. The effective flow of workers from sector $i$ to sector $j$ is therefore given by
\begin{equation}
\Psi^t_{i,j} = \psi^t_{i,j} \min
\left( \frac{|\Delta\ell^t_{h,j}|} {\sum_k \psi^t_{k,j}}, 1 \right).
\end{equation}
\end{enumerate}
Hence, the scaling factor ensures that the total effective inflow into industry $j$ does not exceed the magnitude of its negative labour gap ($|\Delta\ell^t_{h,j}|$). The magnitude of labour flows from sector $i$ to sector $j$ depends on the negative labour gap in destination $j$, the positive labour gap in origin $i$ ($\Delta \ell^t_{h,i}$), the wage differential $p^t_{h,j}-p^t_{h,i}$ and the similarity between sector $i$ and $j,$ $s_{i,j}$. It is straightforward to verify that labour supply dynamics are at a rest point when wages are constant across sectors. Otherwise, labour reallocation tends to reduce wage differentials by decreasing (increasing) labour supply in sectors where wages are below (above) the average wage for the corresponding worker type. Thus, given a change in wages, worker reallocation depends on the parameter $\tau_h$ and on the transferability of skills, measured by the similarity matrix $(s_{i,j})_{i,j \in \mathcal{F}}.$

\section{Data and industry classifications}
\label{sec:digital_class}\label{SM-sec:digital_class}
The model uses the 2022 EXIOBASE benchmark with 163 industries in each of 49 regions. Table~\ref{tab:S_industries} reproduces the industry classifications supplied for this experiment. Production nests, household nests, and broad economic sectors are distinct classifications. In particular, transport and waste are household categories but are aggregated within the six economic sectors used to display the results.

\subsection*{Digital suppliers and exposure}
A digital-input flag of one identifies the six industries receiving the productivity shock. Exposure is a separate classification: 49 industries are designated Tier 2 (T2) and 114 Tier 1 (T1). Under the high-embeddedness scenario, only T2 industries combine their digital inputs with labour and capital in the value-added production nest. Under the low-embeddedness scenario, every industry retains the standard production partition shown in the table. T1 and T2 are the implemented scenario designations.

\vspace{20pt}
\begingroup\scriptsize
\setlength{\tabcolsep}{2pt}
\begin{longtable}{P{0.7cm}P{6cm}P{1.2cm}P{2cm}P{2cm}P{2cm}P{1.2cm}}
\caption{Industry classifications.}\label{tab:S_industries}\label{SM-tab:S_industries}\\
\toprule Code & Industry & Digital input & Firm's input nest & Household's item nest & Economic sector & Industry exposure \\\midrule
\endfirsthead
\toprule Code & Industry & Digital input & Firm's input nest & Household's item nest & Economic sector & Industry exposure \\\midrule
\endhead
\bottomrule\endfoot
1001 & Cultivation of paddy rice & 0 & Agric./forestry & Agri./food & Agri. \& Food & T1 \\
1002 & Cultivation of wheat & 0 & Agric./forestry & Agri./food & Agri. \& Food & T1 \\
1003 & Cultivation of cereal grains nec & 0 & Agric./forestry & Agri./food & Agri. \& Food & T1 \\
1004 & Cultivation of vegetables, fruit, nuts & 0 & Agric./forestry & Agri./food & Agri. \& Food & T1 \\
1005 & Cultivation of oil seeds & 0 & Agric./forestry & Agri./food & Agri. \& Food & T1 \\
1006 & Cultivation of sugar cane, sugar beet & 0 & Agric./forestry & Agri./food & Agri. \& Food & T1 \\
1007 & Cultivation of plant-based fibers & 0 & Agric./forestry & Agri./food & Agri. \& Food & T1 \\
1008 & Cultivation of crops nec & 0 & Agric./forestry & Agri./food & Agri. \& Food & T1 \\
1009 & Cattle farming & 0 & Agric./forestry & Agri./food & Agri. \& Food & T1 \\
1010 & Pigs farming & 0 & Agric./forestry & Agri./food & Agri. \& Food & T1 \\
1011 & Poultry farming & 0 & Agric./forestry & Agri./food & Agri. \& Food & T1 \\
1012 & Meat animals nec & 0 & Agric./forestry & Agri./food & Agri. \& Food & T1 \\
1013 & Animal products nec & 0 & Agric./forestry & Agri./food & Agri. \& Food & T1 \\
1014 & Raw milk & 0 & Agric./forestry & Agri./food & Agri. \& Food & T1 \\
1015 & Wool, silk-worm cocoons & 0 & Agric./forestry & Agri./food & Agri. \& Food & T1 \\
1016 & Manure treatment (conventional), storage and land application & 0 & Agric./forestry & Agri./food & Agri. \& Food & T1 \\
1017 & Manure treatment (biogas), storage and land application & 0 & Agric./forestry & Agri./food & Agri. \& Food & T1 \\
1018 & Forestry, logging and related service activities (02) & 0 & Agric./forestry & Agri./food & Agri. \& Food & T1 \\
1019 & Fishing, operating of fish hatcheries and fish farms; service activities incidental to fishing (05) & 0 & Agric./forestry & Agri./food & Agri. \& Food & T1 \\
1020 & Mining of coal and lignite; extraction of peat (10) & 0 & Fossil inputs & Mining & Mining & T1 \\
1021 & Extraction of crude petroleum and services related to crude oil extraction, excluding surveying & 0 & Fossil inputs & Mining & Mining & T1 \\
1022 & Extraction of natural gas and services related to natural gas extraction, excluding surveying & 0 & Fossil inputs & Mining & Mining & T1 \\
1023 & Extraction, liquefaction, and regasification of other petroleum and gaseous materials & 0 & Fossil inputs & Mining & Mining & T1 \\
1024 & Mining of uranium and thorium ores (12) & 0 & Mining & Mining & Mining & T1 \\
1025 & Mining of iron ores & 0 & Mining & Mining & Mining & T1 \\
1026 & Mining of copper ores and concentrates & 0 & Mining & Mining & Mining & T1 \\
1027 & Mining of nickel ores and concentrates & 0 & Mining & Mining & Mining & T1 \\
1028 & Mining of aluminium ores and concentrates & 0 & Mining & Mining & Mining & T1 \\
1029 & Mining of precious metal ores and concentrates & 0 & Mining & Mining & Mining & T1 \\
1030 & Mining of lead, zinc and tin ores and concentrates & 0 & Mining & Mining & Mining & T1 \\
1031 & Mining of other non-ferrous metal ores and concentrates & 0 & Mining & Mining & Mining & T1 \\
1032 & Quarrying of stone & 0 & Mining & Mining & Mining & T1 \\
1033 & Quarrying of sand and clay & 0 & Mining & Mining & Mining & T1 \\
1034 & Mining of chemical and fertilizer minerals, production of salt, other mining and quarrying n.e.c. & 0 & Mining & Mining & Mining & T1 \\
1035 & Processing of meat cattle & 0 & Agric./forestry & Agri./food & Agri. \& Food & T1 \\
1036 & Processing of meat pigs & 0 & Agric./forestry & Agri./food & Agri. \& Food & T1 \\
1037 & Processing of meat poultry & 0 & Agric./forestry & Agri./food & Agri. \& Food & T1 \\
1038 & Production of meat products nec & 0 & Agric./forestry & Agri./food & Agri. \& Food & T1 \\
1039 & Processing vegetable oils and fats & 0 & Manufacturing & Agri./food & Agri. \& Food & T1 \\
1040 & Processing of dairy products & 0 & Agric./forestry & Agri./food & Agri. \& Food & T1 \\
1041 & Processed rice & 0 & Agric./forestry & Agri./food & Agri. \& Food & T1 \\
1042 & Sugar refining & 0 & Manufacturing & Agri./food & Agri. \& Food & T1 \\
1043 & Processing of Food products nec & 0 & Agric./forestry & Agri./food & Agri. \& Food & T1 \\
1044 & Manufacture of beverages & 0 & Agric./forestry & Agri./food & Agri. \& Food & T1 \\
1045 & Manufacture of fish products & 0 & Agric./forestry & Agri./food & Agri. \& Food & T1 \\
1046 & Manufacture of tobacco products (16) & 0 & Agric./forestry & Agri./food & Agri. \& Food & T1 \\
1047 & Manufacture of textiles (17) & 0 & Manufacturing & Manufacturing & Manufacturing & T1 \\
1048 & Manufacture of wearing apparel; dressing and dyeing of fur (18) & 0 & Manufacturing & Manufacturing & Manufacturing & T1 \\
1049 & Tanning and dressing of leather; manufacture of luggage, handbags, saddlery, harness and footwear (19) & 0 & Manufacturing & Manufacturing & Manufacturing & T1 \\
1050 & Manufacture of wood and of products of wood and cork, except furniture; manufacture of articles of straw and plaiting materials (20) & 0 & Manufacturing & Manufacturing & Manufacturing & T2 \\
1051 & Re-processing of secondary wood material into new wood material & 0 & Manufacturing & Manufacturing & Manufacturing & T2 \\
1052 & Pulp & 0 & Manufacturing & Manufacturing & Manufacturing & T2 \\
1053 & Re-processing of secondary paper into new pulp & 0 & Manufacturing & Manufacturing & Manufacturing & T2 \\
1054 & Paper & 0 & Manufacturing & Manufacturing & Manufacturing & T2 \\
1055 & Publishing, printing and reproduction of recorded media (22) & 0 & Manufacturing & Manufacturing & Manufacturing & T2 \\
1056 & Manufacture of coke oven products & 0 & Energy & Energy & Energy & T1 \\
1057 & Petroleum Refinery & 0 & Energy & Energy & Energy & T1 \\
1058 & Processing of nuclear fuel & 0 & Energy & Energy & Energy & T1 \\
1059 & Plastics, basic & 0 & Manufacturing & Manufacturing & Manufacturing & T1 \\
1060 & Re-processing of secondary plastic into new plastic & 0 & Manufacturing & Manufacturing & Manufacturing & T1 \\
1061 & N-fertiliser & 0 & Manufacturing & Manufacturing & Manufacturing & T1 \\
1062 & P- and other fertiliser & 0 & Manufacturing & Manufacturing & Manufacturing & T1 \\
1063 & Chemicals nec & 0 & Manufacturing & Manufacturing & Manufacturing & T1 \\
1064 & Manufacture of rubber and plastic products (25) & 0 & Manufacturing & Manufacturing & Manufacturing & T1 \\
1065 & Manufacture of glass and glass products & 0 & Manufacturing & Manufacturing & Manufacturing & T1 \\
1066 & Re-processing of secondary glass into new glass & 0 & Manufacturing & Manufacturing & Manufacturing & T1 \\
1067 & Manufacture of ceramic goods & 0 & Manufacturing & Manufacturing & Manufacturing & T1 \\
1068 & Manufacture of bricks, tiles and construction products, in baked clay & 0 & Manufacturing & Manufacturing & Manufacturing & T1 \\
1069 & Manufacture of cement, lime and plaster & 0 & Manufacturing & Manufacturing & Manufacturing & T1 \\
1070 & Re-processing of ash into clinker & 0 & Manufacturing & Manufacturing & Manufacturing & T1 \\
1071 & Manufacture of other non-metallic mineral products n.e.c. & 0 & Manufacturing & Manufacturing & Manufacturing & T1 \\
1072 & Manufacture of basic iron and steel and of ferro-alloys and first products thereof & 0 & Manufacturing & Manufacturing & Manufacturing & T1 \\
1073 & Re-processing of secondary steel into new steel & 0 & Manufacturing & Manufacturing & Manufacturing & T1 \\
1074 & Precious metals production & 0 & Manufacturing & Manufacturing & Manufacturing & T1 \\
1075 & Re-processing of secondary preciuos metals into new preciuos metals & 0 & Manufacturing & Manufacturing & Manufacturing & T1 \\
1076 & Aluminium production & 0 & Manufacturing & Manufacturing & Manufacturing & T1 \\
1077 & Re-processing of secondary aluminium into new aluminium & 0 & Manufacturing & Manufacturing & Manufacturing & T1 \\
1078 & Lead, zinc and tin production & 0 & Manufacturing & Manufacturing & Manufacturing & T1 \\
1079 & Re-processing of secondary lead into new lead, zinc and tin & 0 & Manufacturing & Manufacturing & Manufacturing & T1 \\
1080 & Copper production & 0 & Manufacturing & Manufacturing & Manufacturing & T1 \\
1081 & Re-processing of secondary copper into new copper & 0 & Manufacturing & Manufacturing & Manufacturing & T1 \\
1082 & Other non-ferrous metal production & 0 & Manufacturing & Manufacturing & Manufacturing & T1 \\
1083 & Re-processing of secondary other non-ferrous metals into new other non-ferrous metals & 0 & Manufacturing & Manufacturing & Manufacturing & T1 \\
1084 & Casting of metals & 0 & Manufacturing & Manufacturing & Manufacturing & T1 \\
1085 & Manufacture of fabricated metal products, except machinery and equipment (28) & 0 & Manufacturing & Manufacturing & Manufacturing & T1 \\
1086 & Manufacture of machinery and equipment n.e.c. (29) & 0 & Manufacturing & Manufacturing & Manufacturing & T2 \\
1087 & Manufacture of office machinery and computers (30) & 1 & Digital & Digital & Digital Ind. & T2 \\
1088 & Manufacture of electrical machinery and apparatus n.e.c. (31) & 1 & Digital & Digital & Digital Ind. & T2 \\
1089 & Manufacture of radio, television and communication equipment and apparatus (32) & 1 & Digital & Digital & Digital Ind. & T2 \\
1090 & Manufacture of medical, precision and optical instruments, watches and clocks (33) & 1 & Digital & Digital & Digital Ind. & T2 \\
1091 & Manufacture of motor vehicles, trailers and semi-trailers (34) & 0 & Manufacturing & Manufacturing & Manufacturing & T2 \\
1092 & Manufacture of other transport equipment (35) & 0 & Manufacturing & Manufacturing & Manufacturing & T2 \\
1093 & Manufacture of furniture; manufacturing n.e.c. (36) & 0 & Manufacturing & Manufacturing & Manufacturing & T2 \\
1094 & Recycling of waste and scrap & 0 & Manufacturing & Manufacturing & Manufacturing & T1 \\
1095 & Recycling of bottles by direct reuse & 0 & Manufacturing & Manufacturing & Manufacturing & T1 \\
1096 & Production of electricity by coal & 0 & Energy & Energy & Energy & T2 \\
1097 & Production of electricity by gas & 0 & Energy & Energy & Energy & T2 \\
1098 & Production of electricity by nuclear & 0 & Energy & Energy & Energy & T2 \\
1099 & Production of electricity by hydro & 0 & Energy & Energy & Energy & T2 \\
1100 & Production of electricity by wind & 0 & Energy & Energy & Energy & T2 \\
1101 & Production of electricity by petroleum and other oil derivatives & 0 & Energy & Energy & Energy & T2 \\
1102 & Production of electricity by biomass and waste & 0 & Energy & Energy & Energy & T2 \\
1103 & Production of electricity by solar photovoltaic & 0 & Energy & Energy & Energy & T2 \\
1104 & Production of electricity by solar thermal & 0 & Energy & Energy & Energy & T2 \\
1105 & Production of electricity by tide, wave, ocean & 0 & Energy & Energy & Energy & T2 \\
1106 & Production of electricity by Geothermal & 0 & Energy & Energy & Energy & T2 \\
1107 & Production of electricity nec & 0 & Energy & Energy & Energy & T2 \\
1108 & Transmission of electricity & 0 & Manufacturing & Services & Services & T2 \\
1109 & Distribution and trade of electricity & 0 & Services & Services & Services & T2 \\
1110 & Manufacture of gas; distribution of gaseous fuels through mains & 0 & Energy & Energy & Energy & T2 \\
1111 & Steam and hot water supply & 0 & Energy & Energy & Energy & T2 \\
1112 & Collection, purification and distribution of water (41) & 0 & Services & Services & Services & T1 \\
1113 & Construction (45) & 0 & Services & Services & Services & T1 \\
1114 & Re-processing of secondary construction material into aggregates & 0 & Services & Services & Services & T1 \\
1115 & Sale, maintenance, repair of motor vehicles, motor vehicles parts, motorcycles, motor cycles parts and accessoiries & 0 & Services & Services & Services & T2 \\
1116 & Retail sale of automotive fuel & 0 & Energy & Energy & Energy & T2 \\
1117 & Wholesale trade and commission trade, except of motor vehicles and motorcycles (51) & 0 & Services & Services & Services & T2 \\
1118 & Retail trade, except of motor vehicles and motorcycles; repair of personal and household goods (52) & 0 & Services & Services & Services & T2 \\
1119 & Hotels and restaurants (55) & 0 & Services & Services & Services & T1 \\
1120 & Transport via railways & 0 & Services & Transport & Services & T1 \\
1121 & Other land transport & 0 & Services & Transport & Services & T1 \\
1122 & Transport via pipelines & 0 & Services & Manufacturing & Services & T1 \\
1123 & Sea and coastal water transport & 0 & Services & Transport & Services & T1 \\
1124 & Inland water transport & 0 & Services & Transport & Services & T1 \\
1125 & Air transport (62) & 0 & Services & Transport & Services & T1 \\
1126 & Supporting and auxiliary transport activities; activities of travel agencies (63) & 0 & Services & Services & Services & T1 \\
1127 & Post and telecommunications (64) & 1 & Digital & Digital & Digital Ind. & T2 \\
1128 & Financial intermediation, except insurance and pension funding (65) & 0 & Services & Services & Services & T2 \\
1129 & Insurance and pension funding, except compulsory social security (66) & 0 & Services & Services & Services & T2 \\
1130 & Activities auxiliary to financial intermediation (67) & 0 & Services & Services & Services & T2 \\
1131 & Real estate activities (70) & 0 & Services & Services & Services & T1 \\
1132 & Renting of machinery and equipment without operator and of personal and household goods (71) & 0 & Services & Services & Services & T2 \\
1133 & Computer and related activities (72) & 1 & Digital & Digital & Digital Ind. & T2 \\
1134 & Research and development (73) & 0 & Services & Services & Services & T2 \\
1135 & Other business activities (74) & 0 & Services & Services & Services & T2 \\
1136 & Public administration and defence; compulsory social security (75) & 0 & Services & Services & Services & T2 \\
1137 & Education (80) & 0 & Services & Services & Services & T2 \\
1138 & Health and social work (85) & 0 & Services & Services & Services & T2 \\
1139 & Incineration of waste: Food & 0 & Services & Manufacturing & Manufacturing & T1 \\
1140 & Incineration of waste: Paper & 0 & Services & Manufacturing & Manufacturing & T1 \\
1141 & Incineration of waste: Plastic & 0 & Services & Manufacturing & Manufacturing & T1 \\
1142 & Incineration of waste: Metals and Inert materials & 0 & Services & Manufacturing & Manufacturing & T1 \\
1143 & Incineration of waste: Textiles & 0 & Services & Manufacturing & Manufacturing & T1 \\
1144 & Incineration of waste: Wood & 0 & Services & Manufacturing & Manufacturing & T1 \\
1145 & Incineration of waste: Oil/Hazardous waste & 0 & Services & Manufacturing & Manufacturing & T1 \\
1146 & Biogasification of food waste, incl. land application & 0 & Services & Waste & Services & T2 \\
1147 & Biogasification of paper, incl. land application & 0 & Services & Waste & Services & T2 \\
1148 & Biogasification of sewage slugde, incl. land application & 0 & Services & Waste & Services & T2 \\
1149 & Composting of food waste, incl. land application & 0 & Services & Waste & Services & T1 \\
1150 & Composting of paper and wood, incl. land application & 0 & Services & Waste & Services & T1 \\
1151 & Waste water treatment, food & 0 & Services & Waste & Services & T1 \\
1152 & Waste water treatment, other & 0 & Services & Waste & Services & T1 \\
1153 & Landfill of waste: Food & 0 & Services & Waste & Services & T1 \\
1154 & Landfill of waste: Paper & 0 & Services & Waste & Services & T1 \\
1155 & Landfill of waste: Plastic & 0 & Services & Waste & Services & T1 \\
1156 & Landfill of waste: Inert/metal/hazardous & 0 & Services & Waste & Services & T1 \\
1157 & Landfill of waste: Textiles & 0 & Services & Waste & Services & T1 \\
1158 & Landfill of waste: Wood & 0 & Services & Waste & Services & T1 \\
1159 & Activities of membership organisation n.e.c. (91) & 0 & Services & Services & Services & T1 \\
1160 & Recreational, cultural and sporting activities (92) & 0 & Services & Services & Services & T2 \\
1161 & Other service activities (93) & 0 & Services & Services & Services & T1 \\
1162 & Private households with employed persons (95) & 0 & Services & Services & Services & T1 \\
1163 & Extra-territorial organizations and bodies & 0 & Services & Services & Services & T1 \\
\end{longtable}
\endgroup

\section{Model calibration and technical details}
\label{sec:S_model_calibration}\label{SM-sec:S_model_calibration}

\subsection*{Benchmark equilibrium and expenditure shares}

This section describes the calibration of the nested CES production and utility functions at the benchmark equilibrium and the out-of-equilibrium allocation of inputs and consumption.

The notation follows the model definition: $i \in \F$ denotes a producing firm, $j \in \M$ denotes an input good, $h \in \H$ denotes a household, and $s$ denotes a nest in the corresponding partition of goods.

The calibration is performed on the 2022 benchmark input--output equilibrium. Let $p_j$ denote the benchmark price of good $j$ and let $x_{ji}$ denote the quantity of good $j$ used by firm $i$. Total nominal expenditure by firm $i$ is
\begin{equation}
    E_i = \sum_{j \in \M} p_j x_{ji}.
\end{equation}
For a nest $s \in S_i$, the nominal expenditure of firm $i$ on that nest is
\begin{equation}
    E_{s,i} = \sum_{j \in s} p_j x_{ji}.
\end{equation}
The corresponding expenditure shares are
\begin{equation}
    \omega_{j,i} = \frac{p_j x_{ji}}{E_i},
    \qquad
    \Omega_{s,i} = \frac{E_{s,i}}{E_i}.
\end{equation}
The same notation is used for households, replacing the firm index $i$ with the household index $h$ and using observed benchmark consumption $x_{jh}$.

\subsection*{Calibration of nested CES production and utility functions}

For each firm $i \in \F$, production is represented by the nested CES technology
\begin{equation}
    f_i(x_i) = \Lambda_i \phi_i(x_i),
\end{equation}
with
\begin{equation}
    \phi_i(x_i)=
    \left[
      \sum_{s \in S_i}
      \left(
        \sum_{j \in s} \beta_{ji} x_{ji}^{\rho_s}
      \right)^{\theta / \rho_s}
    \right]^{1/\theta}.
\end{equation}
At the benchmark equilibrium, productivity is normalised to $\Lambda_i=1$. The parameters $\beta_{ji}$ are calibrated so that the observed input vector reproduces benchmark output $y_i$ and benchmark expenditure shares.

Therefore, after solving the first-order conditions, the $\beta$-parameters for the production function of firm $i$ are
\begin{equation}
    \label{eq:beta-production-calibration}
    \beta_{ji} = 
    \frac{p_{j}x_{ji}}{\sum_k p_{k}x_{ki}} 
    \left [ \frac{\sum_{k \in s} p_{k}x_{k,i}}{\sum_k p_{k}x_{k,i}} \right]^{\frac{\rho_s-\theta}{\theta}}
    \left (\frac{y_i}{x_{ji}}\right)^{\rho_s} =
    \omega_{j,i}
    \Omega_{s,i}^{\frac{\rho_s-\theta}{\theta}}
    \left(\frac{y_i}{x_{ji}}\right)^{\rho_s}\;,
\end{equation}
where $j$ denotes an input belonging to nest $s$, and $s \in S_i$ indexes the set of production nests of firm $i$. The calibration procedure to maximise household utility follows the same logic. 

The elasticity associated with the top-level aggregator is
\begin{equation}
    \sigma = \frac{1}{1-\theta} \;,
\end{equation}
while the elasticity inside nest $s$ is
\begin{equation}
    \sigma_s = \frac{1}{1-\rho_s}.
\end{equation}
These elasticities are defined whenever $\theta \neq 1$ and $\rho_s \neq 1$.

The calibrated matrices $\beta_{ji}$ and $\beta_{jh}$ are computed only for observed links in the benchmark input--output table. Non-observed links retain zero weight. Labour and capital services enter the value-added nest. Under high-embeddedness, digital inputs enter this same nest for T2 industries. The calibration formulas apply conditional on the scenario-specific partition $S_i$, so both structures reproduce the same initial input--output observations.

\subsection*{Weight parametrisation}

The above optimisation problem corresponds to the benchmark production problem: given prices and a nominal input budget, the firm chooses the input bundle that maximises production. We now consider the corresponding expenditure-allocation problem after prices change, taking the calibrated parameters $\beta_{ji}$, $\rho_s$, and $\theta$ as given.

For firm $i$, define the nest-specific aggregator
\begin{equation}
    G_{s,i}(x_i)=
    \left(
        \sum_{j \in s}\beta_{ji}x_{ji}^{\rho_s}
    \right)^{1/\rho_s},
    \qquad s \in S_i.
\end{equation}

The production aggregator is then
\begin{equation}
    \phi_i(x_i)=
    \left(
        \sum_{s \in S_i}G_{s,i}(x_i)^\theta
    \right)^{1/\theta}.
\end{equation}

Once the production function has been calibrated, the parameters $\beta_{ji}$, $\rho_s$, and $\theta$ remain fixed throughout the simulations. Out of equilibrium, firms do not re-optimise these parameters. Instead, they adjust their input bundles in response to changes in relative prices while preserving the calibrated production technology.

To characterise these adjustments, it is convenient to reformulate the firm's problem as a cost-minimisation problem. Given a target level of production, firms choose the least-cost combination of inputs subject to the nested CES production structure. The solution yields the optimal expenditure shares associated with each production nest and, subsequently, the optimal allocation of expenditure across individual inputs within each nest.

\subsection*{Cost minimisation and out-of-equilibrium input allocation}

For a given target output level normalised to one, the firm solves
\begin{equation}
\min_{\{x_{ji}\}}
C_i=\sum_{j\in\M} p_j x_{ji}
\qquad
\text{s.t.}
\qquad
\phi_i(x_i)=1.
\end{equation}

The optimisation can be solved recursively. First, expenditure is allocated across inputs within each nest. For a given level of the nest-specific aggregate $\bar G_{s,i}$, the corresponding Lagrangian is
\begin{equation}
\mathcal{L}_{s,i}
=
\sum_{j\in s} p_j x_{ji}
+
\lambda_s
\left(
\bar G_{s,i}-G_{s,i}
\right).
\end{equation}

The solution yields the nest-specific price index
\begin{equation}
\Pi_{s,i}
=
\left(
\sum_{j\in s}
p_j^{\,1-\sigma_s}
\beta_{ji}^{\,\sigma_s}
\right)^{\frac{1}{1-\sigma_s}},
\end{equation}
where
\begin{equation}
\sigma_s=\frac{1}{1-\rho_s}.
\end{equation}
The nest-specific price indices $\Pi_{s,i}$ represent the minimum unit cost of obtaining one unit of the composite input associated with nest $s$.

In a subsequent step, expenditure is allocated across nests. The corresponding optimisation problem is
\begin{equation}
\mathcal{L}_i
=
\sum_{s\in S_i}
\Pi_{s,i}G_{s,i}
+
\lambda
\left(
1-\phi_i
\right).
\end{equation}

Solving this problem yields the unit-cost function
\begin{equation}
C_i
=
\left[
\sum_{s\in S_i}
\Pi_{s,i}^{\,1-\sigma}
\right]^{\frac{1}{1-\sigma}}
=
\left[
\sum_{s\in S_i}
\left(
\sum_{j\in s}
p_j^{\,1-\sigma_s}
\beta_{ji}^{\,\sigma_s}
\right)^{\frac{1-\sigma}{1-\sigma_s}}
\right]^{\frac{1}{1-\sigma}} \; .
\end{equation}
The unit-cost function summarises the minimum expenditure required to produce one unit of output given the prevailing vector of input prices. Applying Shephard's lemma yields the optimal conditional demand for each input,
\begin{equation}
x_{ji}^{*}
=
\frac{\partial C_i}{\partial p_j},
\end{equation}
which determines the expenditure shares and input coefficients used in the out-of-equilibrium dynamics.

\subsection*{Simulation parameters}

The CES weights are calibrated separately for each production structure to reproduce the same initial expenditure shares. Table~\ref{tb:sim_parameters} reports the parameters used in the benchmark simulations, together with the ranges considered in the sensitivity analysis. The latter explores alternative magnitudes of digital-supplier productivity improvements ($g_D$) and value-added substitution elasticities ($\sigma_{\mathrm{VA}}$), while keeping the remaining parameters fixed at their benchmark values.
\begin{table}[t]
\begin{minipage}{0.94\linewidth}
\centering
\caption{Main simulation parameters and sensitivity analysis.}
\label{tb:sim_parameters}\label{SM-tb:sim_parameters}
\footnotesize
\setlength{\tabcolsep}{4pt}
\renewcommand{\arraystretch}{1.08}
\begin{tabular}{@{}p{0.18\linewidth}p{0.5\linewidth}p{0.32\linewidth}@{}}
\toprule
Parameter & Description & Value \\
\midrule
\multicolumn{3}{@{}l}{\textit{Simulation and adjustment}} \\
$g_D$ & Cumulative digital-supplier TFP increase & $0.35$ \\
$T$ & Simulation horizon & 320 iterations (1+10+5 years) \\
$\Delta t$ & Length of iterations per year & 20 \\
$t'$ & Pre-shock period & 20 iterations (1 year) \\
$T_g$ & Digital TFP growth period & 200 iterations (10 years) \\
$T-t'-T_g$ & Constant-TFP period & 100 iterations (5 years) \\
$\tau_p$, $\tau_w$ & Price/wage and expenditure-share adjustment speeds & $0.5$; $0.5$ \\
$\chi$ & Target inventory-to-output ratio & $5$ \\
\midrule
\multicolumn{3}{@{}l}{\textit{CES substitution elasticities}} \\
$\sigma$ & Upper-level production elasticity & $0.33$ \\
$\sigma_{\mathrm{VA}}$ & Capital--labour and augmented digital--VA nests & $0.80$ \\
$\sigma_{\mathrm{other}}$ & Production within-nest elasticities & $0.50$ \\
$\sigma'$ & Upper-level household elasticity & $0.33$ \\
$\sigma'_{\mathrm{other}}$ & Household within-nest elasticities & $0.50$ \\
\midrule
\multicolumn{3}{@{}l}{\textit{Labour mobility and scenarios}} \\
$\tau_h$ & Mobility speeds: sticky; frictional (baseline); rapid & $0$; $0.015$; $0.99$ \\
$\bar{s}$ & Industry-similarity threshold & $0.20$ \\
Production nesting & Alternative digital-input partitions & Low-embeddedness; High-embeddedness \\
\midrule
\multicolumn{3}{@{}l}{\textit{Sensitivity analysis}} \\
$g_D$ & Cumulative digital-supplier TFP increase & $0.10$--$0.40$ (step $0.05$) \\
$\sigma_{\mathrm{VA}}$ & Capital--labour and augmented digital--VA elasticities & \{$0.50$, $0.67$, $0.83$, $0.95$, $1.11$, $1.25$\} \\
\bottomrule
\end{tabular}
\end{minipage}
\end{table}

\section{Production-network measures}
\label{sec:S_network_measures}\label{SM-sec:S_network_measures}

The Results section uses strength and Katz centrality to describe direct and indirect input linkages, alongside upstreamness and downstreamness indicators of network position. These statistics are transformations of the same input-coefficient network. Agreement between strength and Katz is therefore not independent evidence of a causal mechanism, and none of these measures is a transaction-volume statistic.

Under the standard assumption that intermediate input requirements are proportional to sectoral output, let
\[
\mathbf{Z}_t=(z_{ij,t})_{i,j\in\mathcal F}
\]
denote the matrix of real input coefficients, where $z_{ij,t}=x_{ij,t}/y_{j,t}$ measures input supplied by firm $i$ per unit of output produced by firm $j$. Suppliers occupy rows and users occupy columns. Both inputs and output are valued at constant 2022 prices. The coefficients can change as firms adjust their input combinations in response to relative prices, but exclude the mechanical revaluation of unchanged quantities at current prices. The matrix $\mathbf{Z}_t \in \mathbb{R}^{n\times n}$ describes the firm-to-firm component of the production network; feasible supplier links remain fixed.

The input--output accounting identity of the production system can then be written as
\begin{equation}
    Y_t=\mathbf{Z}_tY_t+F_t.
\end{equation}
In this expression, $Y_t$ denotes the vector of gross output and $F_t$ the vector of final demand, both expressed in real terms.
The vector $\mathbf{Z}_tY_t$ therefore corresponds to intermediate demand generated by production activities.

We define the production network as the matrix $\mathbf{Z}$, omitting the time subscript whenever no confusion arises. The element $z_{ij}$ measures the intensity of the production linkage from supplier $i$ to customer $j$, and the resulting weighted directed network describes the structure of inter-firm production relationships. For the sectoral results, we first compute network statistics for individual firms and then sum them within each country and broad economic sector.

\subsection*{Strength and Katz centrality}
The out-strength and in-strength of industry $i$ are defined as the sums of its outgoing and incoming links:
\begin{align}    
\mathbf{s}^{out} &= \mathbf{Z} \mathbf{1} \\
\mathbf{s}^{in} &= \mathbf{Z}^{\top} \mathbf{1} \;.
\end{align}
In-strength measures total real intermediate-input requirements per unit of real output. Because the coefficients are expressed in 2022 prices, they are not generally contemporaneous nominal intermediate-expenditure shares and need not remain bounded by one outside the benchmark. A higher value indicates greater direct intermediate-input requirements. In contrast, out-strength is not bounded by unity. It measures the extent to which an industry's output is used as an intermediate input by other industries, relative to those industries' total production. High out-strength identifies sectors that play an important role as direct suppliers within the production network and whose outputs are widely embedded in other industries' production processes.

Forward Katz centrality captures the importance of an industry as a supplier through both direct and indirect production linkages. Industries with high centrality values are connected to sectors that themselves supply many other sectors, implying that their output propagates widely throughout the economy. This is defined as
\begin{equation}
    \mathbf{k}^{forward} = (I - \alpha \mathbf{Z})^{-1} \mathbf{1} \;;
\end{equation}
This expression converges when $\alpha < \frac{1}{\rho(\mathbf{Z})}$, where $\rho(\mathbf{Z})$ denotes the spectral radius of the production-network matrix. 

Backward Katz centrality follows the transpose orientation and summarises weighted upstream input dependencies. It combines the intensity and reach of paths, rather than measuring chain length alone. This is defined as
\begin{equation}
    \mathbf{k}^{backward} = (I - \alpha \mathbf{Z}^{\top})^{-1} \mathbf{1} \;.
\end{equation}

Because we compare network statistics across scenarios, we need a common attenuation parameter $\alpha$, which is defined as:
\begin{equation}
\alpha=\frac{0.5}{\max_{r\in\mathcal R}\rho(\mathbf Z_r)},
\end{equation}
where $\rho(\mathbf Z_r)$ is the spectral radius of network $r$. For the two comparisons, $\mathcal R$ includes the common initial network and both final scenario networks.

\subsection*{Upstreamness and downstreamness}

Following \cite{antras2012measuring} and \cite{bartolucci2025upstreamness}, we also measure \textit{Upstreamness} and \textit{Downstreamness}, two indicators that provide more information about industries' positions within the production network. Unlike strength and centrality measures, which quantify the intensity and systemic importance of production linkages, upstreamness and downstreamness describe where industries are located along the production process, from primary inputs to final demand.

Upstreamness measures a sector's average distance from final demand and can be interpreted as the expected number of production stages remaining before a sector's output reaches final consumers. Industries with high upstreamness are located further away from final demand and typically supply intermediate inputs that undergo multiple transformations before reaching consumers. Examples include mining, basic materials, and energy-producing industries. This is defined as
\begin{equation}\label{upstream}
U = [\mathbf{I} - \mathbf{Y}^{-1}\mathbf{Z}\mathbf{Y}]^{-1}\mathbf{1} \;.
\end{equation}

Downstreamness captures the average distance from primary factors of production, such as labour, capital, and natural resources. It therefore measures how far an industry is located from the origin of production chains. Industries with high downstreamness tend to operate closer to final demand and combine inputs originating from multiple upstream sectors. Typical examples include consumer-oriented manufacturing and service activities. This is defined as
\begin{equation}\label{downstream}
D = [\mathbf{I} - \mathbf{Z}^{\top}]^{-1}\mathbf{1} \;.
\end{equation}
Equations~\eqref{upstream} and \eqref{downstream}, $\mathbf I$ denotes the identity matrix, $\mathbf Y=\mathrm{diag}(Y)$ is the diagonal matrix of sectoral gross output, $\mathbf Y^{-1}$ is the corresponding inverse diagonal matrix, and $\mathbf 1$ is a column vector of ones.

\section{Production-network recomposition in other sectors}
\label{sec:network_other_sectors}\label{SM-sec:network_other_sectors}
\begin{figure}[h!]
\centering
\includegraphics[width=0.99\linewidth]{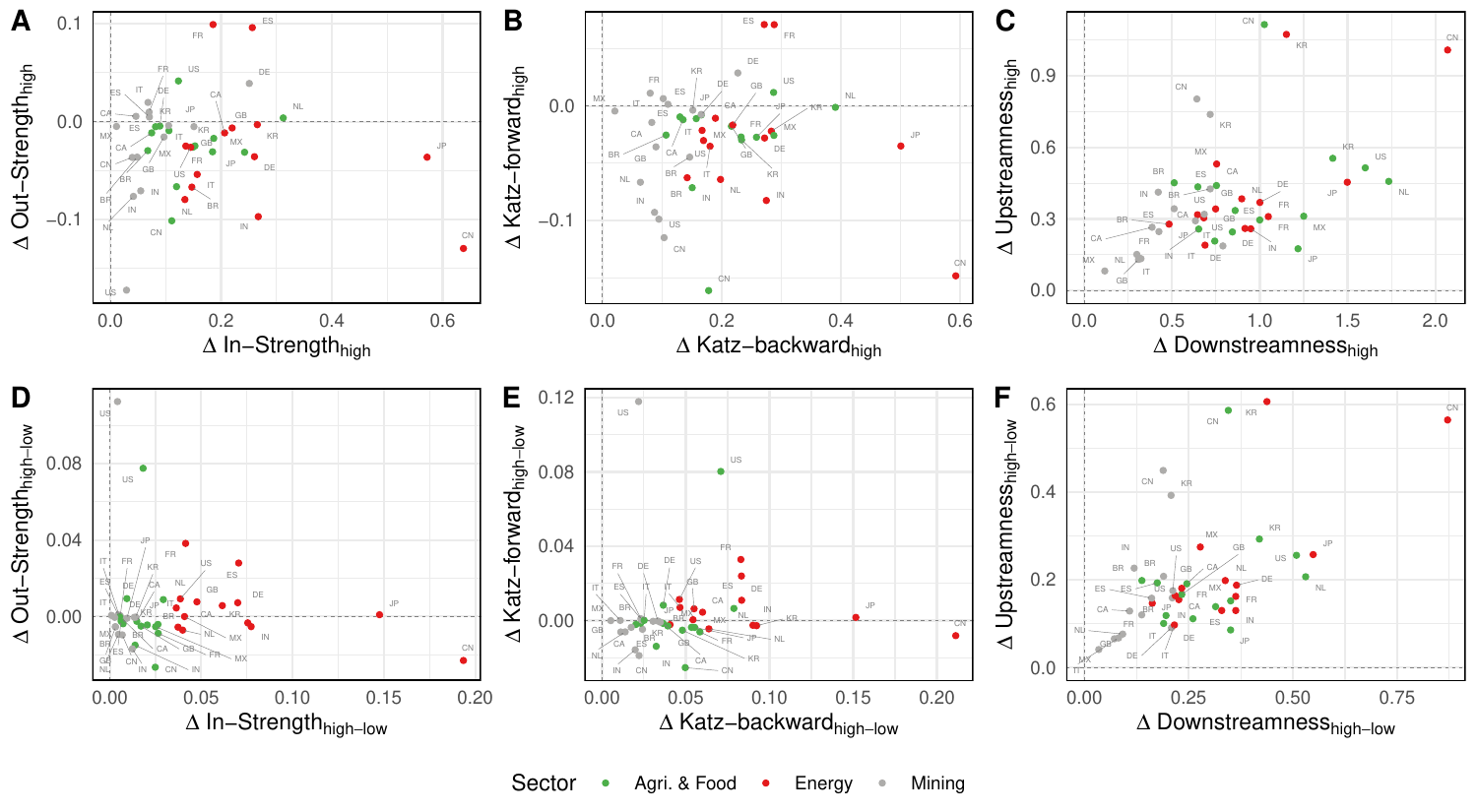}
\caption{Changes in production-network statistics for agriculture and food, energy, and mining. Upper panels (A--C): high-embeddedness scenario relative to $t_0$. Lower panels (D--F): high-embeddedness relative to low-embeddedness at the end of the simulations. Horizontal and vertical axes respectively pair in-strength with out-strength (A, D), backward with forward Katz centrality (B, E), and downstreamness with upstreamness (C, F). Network definitions are provided in Section~\ref{sec:S_network_measures}.}
\label{fig:d_net_delta_sm}
\end{figure}

Figure~\ref{fig:d_net_delta_sm} extends the analysis of production-network recomposition to agriculture and food, energy, and mining. As in the main text, the upper panels report changes under the high-embeddedness scenario relative to $t_0$, while the lower panels compare the high-embeddedness and low-embeddedness scenarios at the end of the simulations.

Relative to the initial benchmark, agriculture and food, energy, and mining generally show stronger backward linkages, with mixed forward responses (Figure~\ref{fig:d_net_delta_sm}, A--B). Energy in France and Spain strengthens both direct and indirect forward linkages, whereas China's energy sector exhibits a pronounced forward decline. These results illustrate that output expansion can coexist with stronger intermediate-input dependence without necessarily increasing an industry's importance as a supplier.

The high-embeddedness versus low-embeddedness comparison reveals further heterogeneity (Figure~\ref{fig:d_net_delta_sm}, D--E). Agriculture and food and mining in the United States display conspicuous positive forward differences. In contrast, China's energy sector combines a large positive backward difference with a negative forward difference. France and Spain retain positive energy forward differences. A positive contrast between scenarios does not necessarily indicate expansion relative to the initial network: US mining, for example, has higher forward statistics than under low embeddedness but lower values than in 2022.

Finally, upstreamness and downstreamness mostly increase in agriculture and food, energy, and mining relative to 2022 (Figure~\ref{fig:d_net_delta_sm}, C). The comparison between scenarios also shows predominantly positive differences in these positional measures, with China's energy sector exhibiting a particularly large downstreamness response (Figure~\ref{fig:d_net_delta_sm}, F).

\bibliographystyle{apalike}
\bibliography{biblio_dig}
\end{document}